\documentclass[10pt,a4paper]{article}

\usepackage[left=2cm,right=2cm,top=2cm,bottom=2cm]{geometry}
\usepackage[english]{babel}
\usepackage{amsmath}
\usepackage{amsfonts}
\usepackage{amssymb}
\usepackage{bbm}
\usepackage{latexsym}
\usepackage{lmodern}
\usepackage{mathrsfs}
\usepackage{slashed}
\usepackage{braket}
\usepackage{tensor}
\usepackage{simplewick}
\usepackage{graphicx}
\usepackage{xcolor}
\usepackage[font=small,labelfont=it]{caption}
\usepackage[font=scriptsize]{caption}
\usepackage[hidelinks]{hyperref}
\hypersetup{
        colorlinks=true,
        linkcolor=blue,
        citecolor=magenta,
        colorlinks = true,
        filecolor=magenta,      
        urlcolor=cyan,
        bookmarksnumbered=true,
        bookmarks=true,
}
\usepackage[nottoc]{tocbibind} 

\newcommand{\al}{\alpha}
\newcommand{\be}{\beta}
\newcommand{\ga}{\gamma}
\newcommand{\de}{\delta}
\newcommand{\eps}{\epsilon}
\newcommand{\ve}{\varepsilon}

\newcommand{\ka}{\kappa}
\newcommand{\la}{\lambda}
\newcommand{\vp}{\varphi}

\newcommand{\om}{\omega}

\newcommand{\sfrac}[2]{{\textstyle\frac{#1}{#2}}}

\newcommand{\pa}{\partial}
\newcommand{\im}{{\mathrm{i}}}

\newcommand{\ep}{{\mathrm{e}}}
\newcommand{\diff}{{\mathrm{d}}}

\newcommand{\beq}{\begin{equation}}
\newcommand{\eeq}{\end{equation}}
\newcommand{\eq}{\end{equation}}
\newcommand{\bea}{\begin{eqnarray}}
\newcommand{\eea}{\end{eqnarray}}
\newcommand{\with}{{\quad{\rm with}\quad}}
\newcommand{\for}{{\quad{\rm for}\quad}}
\renewcommand{\and}{{\quad{\rm and}\quad}}
\newcommand{\und}{{\qquad{\rm and}\qquad}}
\renewcommand{\=}{\ =\ }

\newcommand{\nn}{\nonumber}

\allowdisplaybreaks  
\DeclareMathOperator{\tr}{tr}

\DeclareMathOperator{\Tr}{Tr}
\newcommand{\ie}{{i.e.}}

\newcommand{\Lagr}{ { \mathcal{L} } }
\newcommand{\Delt}{ { \mathcal{M} } }

\newcommand{\wLagr}{\widetilde{\Lagr}}
\newcommand{\wDelt}{\widetilde{\Delt}}

\newcommand{\wR}{\widetilde{R}}
\newcommand{\wY}{\widetilde{Y}}
\newcommand{\wZ}{\widetilde{Z}}
\newcommand{\wF}{\widetilde{F}}
\newcommand{\wA}{\widetilde{A}}
\newcommand{\wla}{\widetilde{\lambda}}
\newcommand{\wde}{\widetilde{\delta}}
\renewcommand{\wp}{{\widetilde{\phi}}}
\newcommand{\wP}{{\widetilde{\phi}}}
\newcommand{\wps}{{\widetilde{\psi}}}
\newcommand{\wro}{{\widetilde{\rho}}}
\newcommand{\wc}{\widetilde{c}}
\newcommand{\wb}{\widetilde{b}}
\newcommand{\ws}{\widetilde{s}}

\advance \textheight by \topskip
\makeatletter
\@addtoreset{equation}{section}

\makeatother
\usepackage{tikz}
\usetikzlibrary{positioning,arrows}
\usetikzlibrary{decorations.pathmorphing}
\usetikzlibrary{decorations.markings}
\tikzset{
particle/.style={thick,draw=black, postaction={decorate},
    decoration={markings,mark=at position .50 with {\arrow[line width=1pt]{>}}  }},
aparticle/.style={thick,draw=black, postaction={decorate},
    decoration={markings,mark=at position .5 with  {\arrow[line width=1pt]{<}}  }},
gluon/.style={decorate, draw=black,
    decoration={coil,aspect=0}},
scalar/.style={thick,dashed,draw=black}
 }

\begin{document}

\begin{titlepage}
\setcounter{page}{0}

\phantom{.}
\vskip 1.5cm

\begin{center}

{\LARGE \bf 
The coupling flow for supergravity
}
\vspace{12mm}

{\Large Federico Arrighi, Saurish Khandelwal \ and \ Olaf Lechtenfeld}
\\[8mm]
\noindent {\em
Institut f\"ur Theoretische Physik and Riemann Center for Geometry and Physics\\
Leibniz Universit\"at Hannover, Appelstrasse 2, 30167 Hannover, Germany}
\vspace{12mm}

\begin{abstract} 
\noindent
Quantum correlation functions in supersymmetric field theories can be encoded in the Nicolai map.
The infinitesimal inverse map provides their response to a change in a coupling constant through a ``flow equation''.
We investigate the flow in the gravitational constant~$\kappa$ for four-dimensional Poincar\'e supergravity via its 
superconformal formulation. By expressing the full superconformal off-shell action as a supervariation we derive a  
``flow operator'' in the effective vierbein theory, whose exponentiation yields the gauge-invariant part of an all-order 
(inverse) Nicolai map for supergravity. Alas, the BRST gauge fixing adds to this functional differential flow 
operator a multiplicative term, which ruins its derivational property and therefore jeopardizes its connection with the 
Nicolai map. We argue and present an example, however, that such multiplicative flow contributions might be 
rewritable as derivational ones with the help of Wick's theorem in the free effective theory where the Nicolai-transformed 
correlators are ultimately evaluated. At least in the Landau gauge, the supergravity flow equation is shown to be regular 
at~$\kappa{=}0$, allowing for a perturbative expansion around Minkowski geometry. Therewith, we have overcome two 
of the three obstacles met in an earlier attempt towards a Nicolai map for Poincar\'e supergravity. Finally, we compare
with the direct on-shell construction of a Nicolai map to leading order in~$\kappa$.
\end{abstract}

\end{center}

\end{titlepage}
%


\section{Introduction: the Nicolai map} \label{sec:intro}

Consider an arbitrary supersymmetric quantum field theory with a field content~$\{\phi\}$ and a coupling constant~$g$.
To realize the supersymmetry, the fields comprise bosons, fermions, and possibly ghost and nonpropagating auxiliary fields.
Imagine integrating out an ``inner'' subset~$\{\phi_i\}$ of these fields, in particular all fermions, ghosts and auxiliaries.
This leaves one with an effective theory of bosonic ``outer'' fields~$\phi_o$ alone, whose action~$S_g$ consists of 
a local classical part~$S_g^{(0)}$ (from putting $\phi_i{=}0$) and a nonlocal quantum part (from integrating out~$\phi_i$),
\begin{equation} \label{Sloops}
S_g[\phi_o] \= S_g^{(0)}[\phi_o]\ +\ \sum_{r=1}^\infty \hbar^r\,S_g^{(r)}[\phi_o]\ ,
\end{equation}
where $S_g^{(r)}$ comprises all Feynman graphs with $r$ loops of~$\phi_i$.
If $\phi_i$ occur only quadratically in the supersymmetric Lagrangian then this expansion truncates at order~$\hbar$.

Among all possible nonlocal bosonic field theories, the one given by~$S_g$ is special since it has a supersymmetric 
origin. What characterizes this theory is the existence of a Nicolai map~\cite{Nic1,Nic2,Nic3,FL,DL1,L1,L2}
\begin{equation}
T_g\,:\ \phi_o\ \mapsto\ \phi_o' = T_g\,\phi_o\ ,
\end{equation}
a generically nonpolynomial and nonlocal field transformation making contact with the free theory at $g{=}0$.
Its defining property is
\begin{equation} \label{Tdef}
\bigl\langle \,Y[\phi_o]\, \bigr\rangle^{\phi_o}_g
\= \bigl\langle \,Y[T_g^{-1}\phi_o]\, \bigr\rangle^{\phi_o}_0\ ,
\end{equation}
which equates the quantum expectation value~$\langle\cdots\rangle_g^{\phi_o}$ 
of an arbitrary functional~$Y$ in the effective theory at coupling~$g$ to the expectation value of the 
(inversely Nicolai-)transformed functional in the free theory. 
As long as supersymmetry is unbroken and spacetime is asymptotically flat, the vacuum functional is independent
of~$g$, and thus $\langle\ldots\rangle_g$ need not be normalized.
Since $T_g$ and its inverse are usually just given 
as formal power series in~$g$ and~$\phi_o$, the above relation effectively reshuffles the perturbative expansion 
of an interacting correlator into the power-series expansion of~$T_g^{-1}\phi_o$.
Since 2020, decisive progress has been made in the construction of Nicolai maps, 
in particular for supersymmetric gauge theories~\cite{ALMNPP,LR1,LR3,Lprague,MN,LR2,LR4}.

If, after eliminating all auxiliary fields, all $\phi_i$ appear only quadratically in the supersymmetric action 
then $T_g\,\phi_o$ and $T_g^{-1}\phi_o$ are classical (independent of~$\hbar$) and expanded in terms of tree graphs.
The final free expectation value on the right-hand side of~\eqref{Tdef} then glues together by free $\phi_o$~propagators 
the trees rooted in each field argument of~$Y$. No $\phi_i$ loops appear.
This is the case for all Wess--Zumino and super Yang--Mills theories (including matter) with a flat target-space metric.
If however the inner fields are self-interacting, then the  expansion~\eqref{Sloops}
does not truncate, and the Nicolai map can no longer be classical but is also a power series in~$\hbar$~\cite{CLR,L3},
\begin{equation} \label{quantmap}
T_g\,\phi_o \= T_g^{(0)}\phi_o \ +\ \sum_{r=1}^\infty \hbar^r\,T_g^{(r)}\phi_o\ ,
\end{equation}
in addition to the expansion in powers of~$g$. Also here,
$r$~denotes the number of ``loop decorations'' in a graphical tree expansion.
This happens in supersymmetric nonlinear sigma models and in supergravity.

Comparing the path integrals in~\eqref{Tdef} after substituting $\phi_o\mapsto T_g\,\phi_o$ on the right-hand side yields
\begin{equation}
\exp\{\tfrac{\im}{\hbar} S_g[\phi_o]\} \= 
\textrm{Det}\bigl(\tfrac{\delta T_g\phi_o}{\delta\phi_o}\bigr)\,\exp\{\tfrac{\im}{\hbar} S_0[T_g\,\phi_o]\}
\end{equation}
which provides the identity
\begin{equation} \label{matchingnew}
S^{(0)}_g[\phi_o]\ +\ \sum_{r\ge1} \hbar^r\,S_g^{(r)}[\phi_o] \=
S^{(0)}_0[T_g\,\phi_o]\ +\ \sum_{r\ge1}\hbar^r\,S^{(r)}_0
\ -\ \im\hbar\,\Tr\ln\sfrac{\delta T_g\phi_o}{\delta\phi_o}
\end{equation}
where $\Tr$ stands for the functional trace.
The terms $S_g^{(r>0)}$ lose their $\phi_o$~dependence at $g{=}0$
and hence the right-hand sum is a constant,
while $S^{(0)}_0$ is a quadratic functional of its argument.
Inserting~\eqref{quantmap} into~\eqref{matchingnew} and separating powers of~$\hbar$
we arrive at an infinite hierarchy of ``Nicolai-map conditions'', one at each loop order. 
The $g$ expansion of the $r$-loop condition begins at order~$g^r$.
The relations at tree and one-loop level are
\begin{equation} \label{matching01}
S^{(0)}_g[\phi_o] \= S^{(0)}_0[T_g^{(0)}\phi_o] \quad\und\quad
S_g^{(1)}[\phi_o] \= S^{(0)}_0[T_g\,\phi_o]\big|_{O(\hbar)} + S^{(1)}_0 - \im\,\Tr\ln\sfrac{\delta T_g^{(0)}\phi_o}{\delta\phi_o} 
\end{equation}
and are known as the free-action and determinant-matching conditions.\footnote{
If $T_g^{(1)}\phi_o$ vanishes then the right-hand side of the $O(\hbar)$ condition stems from the Jacobian determinant 
of a classical Nicolai map, and the left-hand side arises from the $\phi_i$ one-loop determinant.}

The Nicolai-map action extends to arbitrary functionals in a distributive manner,
\begin{equation} \label{distri}
T_g\,Y[\phi_o] \ :=\ Y[T_g\,\phi_o]  \qquad\Leftrightarrow\qquad T_g(\phi_1\,\phi_2) \ :=\ (T_g\,\phi_1)\,(T_g\,\phi_2)\ ,
\end{equation}
for any pair of fields and thus for any functional~$Y$. As a consequence, the infinitesimal map acts as a derivation on~$Y$,
\ie~it obeys the Leibniz rule.
In fact, a successful construction method for the Nicolai map integrates (in~$g$) the infinitesimal change
of $\langle\,Y[\phi]_o\,\rangle^{\phi_o}_g$ under a variation of~$g$. The corresponding coupling flow~\footnote{
Compared with earlier publications on the subject, we have redefined the flow operator~$R$ by 
replacing $\tfrac{\diff}{\diff g}$ with $g\tfrac{\diff}{\diff g}$.}
\begin{equation} \label{flowZ0}
g\tfrac{\diff}{\diff g}\,\bigl\langle\,Y[\phi_o]\,\bigr\rangle_g^{\phi_o} \=
\bigl\langle\,\bigl( g\tfrac{\pa}{\pa g}+R[\phi_o;g] \bigr)\,Y[\phi_o] \,\bigr\rangle_g^{\phi_o}
\end{equation}
is governed by a functional differential ($g$-dependent) operator~$R$ (the ``flow operator'') acting on~$Y$.
Iterating this relation in a Taylor series for $\langle\,Y[\phi_o]\,\rangle_g^{\phi_o}$ around $g{=}0$ yields
\begin{equation} \label{Tinv}
\bigl\langle\,Y[T_g^{-1}\phi_o]\,\bigr\rangle_0^{\phi_o} \= \bigl\langle\,Y[\phi_o]\,\bigr\rangle_g^{\phi_o} \=
\bigl\langle\,\ep^{g(\pa_{g'}+\frac1{g'}R[g'])}\,Y[\phi_o]\big|_{g'=0}\,\bigr\rangle_0^{\phi_o} \=
\bigl\langle\,Y\bigl[\ep^{g(\pa_{g'}+\frac1{g'}R[g'])}\phi_o\big|_{g'=0}\bigr]\,\bigr\rangle_0^{\phi_o}
\end{equation}
and thus a formula for the inverse Nicolai map,
\begin{equation} \label{Tinvformula}
T_g^{-1}[\phi_o]  \= \exp\bigl\{ g\,(\pa_{g'}+\tfrac1{g'}R[\phi_o;g']) \bigr\}\,\phi_o\big|_{g'=0}
\end{equation}
with 
\begin{equation}\label{Rgexp}
R[\phi_o;g] \= g\,R_1[\phi_o]+g^2 R_2[\phi_o]+g^3 R_3[\phi_o] + \ldots
\end{equation}
so that $R[\phi_o;g{=}0]=0$.
A universal formula for $T_g\,\phi_o$ itself can also be derived~\cite{LR1}.

The final frontier for the Nicolai map is locally supersymmetric theories, epitomized by supergravity.
More concretely, four-dimensional Poincar\'e supergravity is our target here,
with the gravitational coupling~$\ka$ playing the role of~$g$.
For an early attempt in this direction, see~\cite{DN}.
Much more recently, we attempted to apply the flow-operator method to this case,
but met with three obstacles~\cite{AKL}. 
Firstly, the minimal off-shell version of Poincar\'e supergravity adds a multiplicative term to the flow equation even before gauge fixing.
Secondly, local supersymmetry does not commute with BRST transformations, giving rise to another multiplicative piece
in the flow equation upon gauge fixing.
Thirdly, a test of regularity at~$\ka{=}0$ (Minkowski space) for setting up a perturbative expansion failed.
This led to a rather negative prospect for the existence of a Nicolai map for supergravity, even though
a brute-force on-shell construction to order~$\ka$ yielded a four-parameter family of maps obeying the free-action condition.
The first two obstructions are fundamental:
The flow equation picks up a multiplicative term~$Z$,
\begin{equation}
\ka\tfrac{\diff}{\diff\ka}\,\bigl\langle\,Y[\phi_o]\,\bigr\rangle_\ka^{\phi_o} \=
\bigl\langle\,\bigl( \ka\tfrac{\pa}{\pa\ka}+R[\phi_o;\ka]+Z[\phi_o;\ka]\bigr)\,Y[\phi_o] \,\bigr\rangle_\ka^{\phi_o}\ ,
\end{equation}
which ruins the derivation property of the flow equation~\eqref{flowZ0}
and invalidates the last equation in~\eqref{Tinv} because the integrated flow is no longer distributive for $Z{\neq}0$.
Therefore, integrating such a flow does not give us an inverse Nicolai map.
The universal formula in~\cite{LR1} is no longer valid either, because its proof also relies on the distributivity.
Nevertheless, the integrated flow equation still allows for an alternative computation of quantum correlators,
\begin{equation} \label{Tinvalt}
\bigl\langle\,Y[\phi_o]\,\bigr\rangle_\ka^{\phi_o} \=
\bigl\langle\,\exp\bigl\{ \ka\,(\pa_{\ka'}+\tfrac1{\ka'}R[\phi_o;\ka']+\tfrac1{\ka'}Z[\phi_o;\ka']) \bigr\}\,
Y[\phi_o]\big|_{\ka'=0}\,\bigr\rangle_0^{\phi_o}\ ,
\end{equation}
partially doing the job of an inverse Nicolai map.
We have proposed in~\cite{AKL} to absorb the multiplicative terms in the $\ka$~flow in a similarity transformation,
yielding a partial (inverse) Nicolai map via
\begin{equation}
\bigl\langle\,Y[\phi_o]\,\bigr\rangle_\ka^{\phi_o} \=
\bigl\langle\,U_\ka[\phi_o]\ Y[T_\ka^{-1}\phi_o]\,\bigr\rangle_0^{\phi_o} \ ,
\end{equation}
where the additional measure factor~$U_\ka$ is constructed perturbatively from $R$ and~$Z$.
The perspective of even a partial Nicolai map for supergravity is important enough to pursue this endeavor.

Our results have motivated a very recent direct attack~\cite{CJL} on the Nicolai-map conditions~\eqref{matchingnew}
for Poincar\'e supergravity without employing the coupling-flow strategy.
In an impressive tour de force, the authors managed to build to order~$\ka^2$ and one loop a family of Nicolai maps
\begin{equation}
T_\ka^{(0)}\phi_o + \hbar\,T_\ka^{(1)}\phi_o \=
\bigl( 1 +  \ka\,T^{(0,1)} + \ka^2 T^{(0,2)} + \hbar\ka\,T^{(1,1)} + \hbar\ka^2 T^{(1,2)} \bigr)\phi_o
\end{equation}
for Poincar\'e supergravity by solving the hierarchy in~\eqref{matchingnew} to order $\ka^2$ and~$\hbar^2$.
Remarkably, they find that the existence of a Nicolai map for a combined graviton-gravitino system requires precisely
the Rarita--Schwinger coupling for the spin-$\tfrac32$ field.

In the present paper, we overcome the first obstruction of~\cite{AKL} by employing the {\it superconformal construction\/} 
of supergravity.
It recovers Poincar\'e supergravity as a gauge-fixed version of a superconformal extension, adding connection fields for dilatation, 
R-symmetry and S-supersymmetry local invariance (part of a Weyl multiplet) as well as a chiral compensator multiplet~$(\phi,\rho,F)$. 
In contrast to Poincar\'e supergravity, the off-shell superconformal action can entirely be written as a supervariation. 
By integrating out all ``conformal'' fields in the quantum theory, we reduce back to Poincar\'e supergravity, and further averaging 
over the gravitini generates a proper (derivational) gauge-invariant part of a flow operator for Poincar\'e supergravity.
On the downside, defining the path integral in the superconformal theory requires a more involved gauge fixing, whose BRST
transformations still do not commute with supersymmetry, again giving rise to a multiplicative contribution to the flow operator.
However, not all hope is lost because multiplicative terms in the flow equation may be converted to derivative ones,
as we shall demonstrate. Because we have not yet been able to achieve such a conversion universally, we cannot claim
to have established the existence of an all-order Nicolai map for supergravity.
Identifying terms missed in the earlier computation~\cite{AKL}, we improve on the consistency check and show that,
up to gauge artefacts, the condition for a regular coupling flow around Minkowski space ($\ka{=}0$) is actually fulfilled, 
removing the third obstacle.
Therefore, we do succeed in setting up a computational scheme for quantum supergravity correlators via
the coupling flow~\eqref{Tinvalt}.
Finally, we make contact with the direct {\it on-shell construction\/}~\cite{AKL,CJL} of a Nicolai map and find agreement 
with the leading-order result of~\cite{CJL} after imposing the Landau gauge. 
This result must still be subjected to a gauge-slice projection if one wants the map to strictly respect the gauge condition.
Reverse-engineering a standard flow operator, however, again requires transmuting a multiplicative to a derivative contribution.

\section{The conformal description of quantum supergravity}\label{sec:conformal}

It is well known that Einstein gravity can be considered as a gauge-fixed version of a conformal theory. The same is true
for supergravity~\cite{KTvN,FvP}, and in the following we make use of it.
The theory is built from a Weyl supermultiplet and a chiral supermultiplet (standard scalar compensator with Weyl weight one), 
containing the fields~\footnote{
We follow the conventions of Freedman and Van Proeyen~\cite{FvP}.}
\begin{equation}
e^a_{\ \mu}\ ,\quad b_\mu\ ,\quad A_\mu\ ,\quad \psi_\mu^\al\ ;\quad \phi\ ,\quad \rho^\al\ ,\quad F
\qquad\textrm{with}\quad \mu,a\in\{0,1,2,3\} \and \al\in\{1,2,3,4\}\ ,
\end{equation}
where the first four are the connections for diffeomorphisms, dilatation, U(1) $R$-symmetry and $Q$-supersymmetry,
respectively, while the connections for local Lorentz, special conformal and $S$-supersymmetry,
$\om_{\mu}^{\ ab}$, $f_\mu^{\ a}$ and $\phi_\mu^\al$, are composite fields determined by the constraints. 
We begin our consideration in the ``nonperturbative scaling'' with respect to the gravitational coupling~$\ka$ 
(of mass dimension~$-1$). Except for the metric or vierbein, we denote these fields by a tilde, and the gravitational 
coupling appears only as an overall factor of~$\tfrac1{\ka^2}$ in the Lagrangian~\cite{FvP}
\begin{equation} \label{confsugra}
\tfrac{1}{\ka^2}\,\wLagr_{\textrm{SUSY}} \= -\tfrac{1}{\ka^2}\,e\,\textrm{Re}\,\bigl[
\bar{\wF}\wF + \bar\wP\,\widetilde\nabla_a\widetilde\nabla^a\wP - \bar\wro\,P_L\widetilde{\slashed\nabla}\wro +
\tfrac{1}{\sqrt{2}}\bar{\wps}_\mu\ga^\mu P_L(\bar{\wF}+\wP\,\widetilde{\slashed\nabla})\wro +
\tfrac12\,\wP\,\bar{\wF}\,\bar{\wps}_\mu\ga^{\mu\nu}P_R\wps_\nu \bigr]\ ,
\end{equation}
with $e=\det e^a_{\ \mu}$ and the chiral projectors $P_{L/R}=\tfrac12(1{\pm}\ga_5)$ acting on Majorana spinors. 
The main complexity resides in the superconformal derivatives
\begin{equation}
\begin{aligned}
\widetilde{\nabla}_a\wP &\= e_a^{\ \mu}\bigl[ 
(\pa_\mu - \widetilde{b}_\mu-\im\,\widetilde{A}_\mu)\,\wP -\tfrac1{\sqrt{2}}\bar{\wps}_\mu P_L\wro \bigr]\ ,\\
P_L \widetilde{\nabla}_a\wro &\= e_a^{\ \mu}P_L \bigl[
(\pa_\mu+\tfrac14\,\widetilde{\om}_\mu^{\ ab}\ga_{ab}-\tfrac32\,\widetilde{b}_\mu+\tfrac{\im}{2}\,\widetilde{A}_\mu)\,\wro
-\tfrac1{\sqrt{2}}(\widetilde{\slashed\nabla}\wP+\wF)\,\wps_\mu-\sqrt{2}\,\wP\,\wp_\mu \bigr]\ ,\\
\widetilde\nabla_a\widetilde\nabla^a\wP &\= e^{a\mu} \bigl[
(\pa_\mu-2\widetilde{b}_\mu-\im\widetilde{A}_\mu)\widetilde\nabla_a\wP 
+\widetilde{\om}_{\mu a}^{\ \ b}\widetilde\nabla_b\wP +2\widetilde{f}_{\mu a}\wP
-\tfrac1{\sqrt{2}}\bar{\wps}_\mu P_L \widetilde\nabla_a\wro +\tfrac1{\sqrt{2}}\bar\wp_\mu\ga_a P_L\wro \bigr]\ ,
\end{aligned}
\end{equation}
where the composite connections~$\widetilde{\om}_\mu^{\ ab}$, $\widetilde{f}_\mu^{\ a}$ and $\wp_\mu^{\ \al}$ are not spelled out here.
(They can be found in~\cite{FvP}.)
In the absence of a cosmological constant, $\wLagr_{\textrm{SUSY}}$ is homogeneously quadratic in the chiral supermultiplet
$(\wP,\wro,\wF)$.\footnote{
This admits the elimination of~$\ka$ by a rescaling of the chiral superfield, exhibiting the no-scale property of the superconformal theory.
Since, however, the scale~$\ka^{-1}$ will be introduced by gauge-fixing, we keep it as a device defining a coupling flow.}
It is known~\cite{FvP} that unfolding the terms in the Lagrangian cancels all $\wF$-dependent terms except for the leading $\bar{\wF}\wF$.
The corresponding action $\widetilde{S}_{\textrm{SUSY}}=\smallint\,\diff^4x\,\wLagr_{\textrm{SUSY}}$ is locally invariant under the 
full superconformal group~SU($2,2|1$), part of which are the global $Q$-supersymmetry transformations
\begin{equation} \label{susy}
\begin{aligned}
\wde e^a_{\ \mu} &\= -\tfrac12\bar\wps_\mu\ga^a\eps\ ,\\
\wde \widetilde{b}_\mu &\= \tfrac12\bar\wp_\mu\eps\ ,\\
\wde \widetilde{A}_\mu &\= \tfrac{\im}{2}\bar\wp_\mu\ga_5\eps\ ,\\
\wde \wps_\mu &\= \bigl(\tfrac14\widetilde\om_{\mu ab}\ga^{ab}+\tfrac12\widetilde{b}_\mu-\tfrac{3\im}{2}\widetilde{A}_\mu\ga_5\bigr)\,\eps\ ,\\
\wde \wP &\= \tfrac{1}{\sqrt{2}}\bar\wro\,P_R\,\eps\ ,\\
\wde \wro &\= \tfrac1{\sqrt{2}}(\widetilde{\slashed\nabla}\wP+\wF)\,\eps\ ,\\
\wde \wF &\= -\tfrac1{\sqrt{2}}\widetilde\nabla_\mu\bar\wro\ga^\mu P_L\,\eps
\end{aligned}
\end{equation}
with a constant Majorana spinor parameter~$\eps$. 
The superspace formulation~\cite{B} tells us that the Lagrangian can be expressed as a supervariation,
\begin{equation}
\wLagr_{\textrm{SUSY}} \= \wde_\al\,\wDelt_\al\ ,
\end{equation}
where we have split off the spinor parameter via $\wde X=\wde_\al X\eps_\al$.
Indeed we have verified this key relation for
\begin{equation} \label{Mtilde}
\wDelt_\al \= e\;\textrm{Re}\,\bigl[ 
\tfrac1{\sqrt{2}} P_L (\bar{\wF} + \wP\,\widetilde{\slashed\nabla})\wro\ +\ \tfrac12\,\wP\,\bar{\wF}\ga^a P_R\wps_a \bigr]_\al\ .
\end{equation}
We stress that this result is an important improvement over  the standard approach to supergravity, where
the attempt to express the Lagrangian as an off-shell supervariation fell short by an auxiliary-field term~\cite{AKL}.

The gauge invariance of the Lagrangian requires a gauge-fixing in the path integral, for which we employ the BRST formalism.
In this framework, one chooses a gauge-fixing function
$\wF_A$ for every gauge parameter, summarily labelled by~$A,B,C,\ldots$, and introduces Faddeev--Popov ghost and
antighost fields as well as Nakanishi--Lautrup auxiliary fields,
\begin{equation} \label{ghosts}
\wc^A\ ,\qquad \wc^{*A} \und \wb^A\ ,
\end{equation}
respectively, to achieve off-shell BRST invariance for the gauge-fixed Lagrangian
$\wLagr_{\textrm{SUSY}}+\wLagr_{\textrm{GF}}$.
The gauge-fixing extension of the supersymmetric Lagrangian reads
\begin{equation}
\wLagr_{\textrm{GF}} \= \tfrac12 \wb^A\,T^{-1}_{AB}\,\wb^B - \wb^A \wF_A - (\ws\,\wF_A)\,\wc^{*A}\ ,
\end{equation}
where $T_{AB}$ is a field-independent invertible matrix containing the gauge-fixing parameters.
The Slavnov variation $\ws$ is just a BRST transformation with the global Grassmann-odd parameter stripped off,
so it involves all the gauge fields in place of the gauge parameters.
The gauge-fixing addition is not supersymmetric and cannot be written as a supervariation. 
Instead, it can be expressed as a Slavnov variation,
\begin{equation}
\wLagr_{\textrm{GF}} \= \ws\,\wDelt_{\textrm{GF}} \quad\with\quad
\wDelt_{\textrm{GF}} \= \wc^{*A}\bigl( \tfrac12 T^{-1}_{AB}\,\wb^B - \wF_A \bigr)\ ,
\end{equation}
and hence the full BRST-invariant action allows for
\begin{equation}
\wLagr_{\textrm{SUSY}} + \wLagr_{\textrm{GF}} \=  \wde_\al\,\wDelt_\al + \ws\,\wDelt_{\textrm{GF}} \ .
\end{equation}
It is important to realize that, for local supersymmetry, $\wde_\al$ and $\ws$ do not anticommute,
\begin{equation} \label{obstruction}
\wde_\al\ws\,\wDelt_\al \neq 0 \und
\{\wde_\al,\ws\}\,\wDelt_{\textrm{GF}} = -\wc^{*A}\,\{\wde_\al,\ws\}\,\wF_A \neq 0\ .
\end{equation}

In conformal supergravity, we need to gauge-fix all seven local symmetries, labelled by the symmetry generators $(P,M,K,D,Y,Q,S)$.
A convenient choice is (remember $g_{\mu\nu}=e^a_{\ \mu}e^b_{\ \nu}\eta_{ab}$)
\begin{equation} \label{gauges}
\widetilde{F}^\nu_P = \pa_\mu(\sqrt{-g}\,g^{\mu\nu})\ ,\quad
\widetilde{F}_M^{ab} = e^a_{\ \mu}\eta^{\mu b}-e^b_{\ \mu}\eta^{\mu a}\ ,\quad
\widetilde{F}_K^\mu  = \widetilde{b}^\mu\ ,\quad
\widetilde{F}_D{+}\im\widetilde{F}_Y = \wP{-}1\ ,\quad
\widetilde{F}_Q^\al  = \ga^\nu \wps_\nu^\al\ ,\quad
\widetilde{F}_S^\al  = \wro^\al\ ,
\end{equation}
and a diagonal gauge-fixing parameter matrix $T^{AB}=\de^{AB}T^A$ with
\begin{equation}
T^{P} = \tfrac16\xi_P\ ,\quad
T^{M} = \xi_M\,\Box\ ,\quad
T^{K} = \xi_K\ ,\quad
T^{D}{+}\im T^{Y} = (\xi_D{+}\im\xi_Y)\Box\ ,\quad
T^{Q} = -\tfrac13 \xi_Q \slashed\pa\ ,\quad
T^{S} = -\xi_S \slashed\pa\ ,
\end{equation}
containing dimensionless gauge parameters~$\xi_A$.
All gauge-fixing functions besides $F_P$ are linear in the fields.
The role of gauge-fixing is twofold. Firstly, in the path integral it removes the overcounting due to the gauge redundancy
and secondly, it serves to render well-defined the propagator of dynamical fields, whose kinetic operator features zero modes
corresponding to gauge freedom. The $(M,K,S)$ gauge fields are composite, and the $(D,Y)$ gauge fields are not propagating.
Since only $(e,\wps,\wP,\wro)$ propagate, we have to pay attention to $(\xi_P,\xi_Q,\xi_D,\xi_Y,\xi_S)$ entering the graviton, 
gravitino, chiral scalar and chiral fermion propagators, respectively. 
We will eventually pick convenient values for $\xi_P$ and $\xi_Q$ and the Landau gauge $\xi_A\to\infty$ for $A=D,Y,S$
to make contact with Poincar\'e supergravity.  For the remaining parameters we choose the Landau gauge from the outset,
\begin{equation}
\xi_M \to \infty \und \xi_K \to \infty \qquad\Leftrightarrow\qquad
e_{ab}=e_{ba} \und \widetilde{b}_\mu=0\ .
\end{equation}
Even for a Landau gauge~$\xi_A{\to}\infty$ it is important to impose~$F_A{=}0$ only after performing any supersymmetry
or Slavnov variation on the fields, and after free-field contractions via free propagators. 
We finish by remarking that, although the gauge-fixed version of the conformal formulation is classically equivalent to
Poincar\'e supergravity, for generic values of~$\xi_A$ they might differ on the quantum level due to the BRST procedure.
Therefore, Landau gauge-fixing of the extra (conformal) redundancies is the best choice for recovering quantum Poincar\'e supergravity.

\section{The flow operator}\label{sec:flow}

We now turn to the $\ka$  flow for the conformal formulation of supergravity.
It concerns a quantum correlation function of a generic functional~$\wY$, which we denote by
\begin{equation}
\bigl\langle\,\wY[\wp_o]\,\bigr\rangle^\wp_\ka \= 
\bigl\langle\bigl\langle\,\wY[\wp_o]\,\bigr\rangle\bigr\rangle^{\wp_o}_\ka \=
\bigl\langle\,\wY[\wp_o]\,\bigr\rangle^{\wp_o}_\ka\ ,
\end{equation}
where the inner bracket describes the inner-field averaging and the last correlator is one in the effective theory.
To obtain the flow equation we derive a functional equation for the above correlator,
\begin{equation} \label{flow}
\begin{aligned}
\ka\,\tfrac{\diff}{\diff \ka}\,\bigl\langle\,\wY[\wp_o]\,\bigr\rangle^{\wp_o}_\ka
&\= \ka\,\tfrac{\diff}{\diff \ka}\,\bigl\langle\bigl\langle\,\wY[\wp_o]\,\bigl\rangle\bigr\rangle^{\wp_o}_\ka \\
&\=\bigl\langle\bigl\langle\,\bigl( \ka\,\tfrac{\pa}{\pa \ka}
\ -\ \tfrac{2\im}{\ka^2}\smallint \wLagr_{\textrm{SUSY}}\ -\ \tfrac{2\im}{\ka^2}\smallint \wLagr_{\textrm{GF}} \bigr)\, 
\wY[\wp_o] \,\bigr\rangle\bigr\rangle^{\wp_o}_\ka\\
&\=\bigl\langle\bigl\langle\,\bigl( \ka\,\tfrac{\pa}{\pa \ka}
\ -\ \tfrac{2\im}{\ka^2}\smallint \wde_\al \wDelt_\al \ -\ \tfrac{2\im}{\ka^2}\smallint \ws \wDelt_{\textrm{GF}} \bigr)\, 
\wY[\wp_o] \,\bigr\rangle\bigr\rangle^{\wp_o}_\ka\\
&\=\bigl\langle\bigl\langle\, \ka\,\tfrac{\pa}{\pa \ka}
\ -\ \tfrac{2\im}{\ka^2}\smallint \wDelt_\al\ \wde_\al\ -\ \tfrac{2\im}{\ka^2}\smallint \wDelt_{\textrm{GF}}\ \ws
\ -\ \tfrac{2\im}{\ka^2}\smallint \wDelt_\al\ \tfrac{\im}{\ka^2}\smallint \wde_\al \wLagr_{\textrm{GF}} \,\bigr\rangle\ 
\wY[\wp_o] \,\bigr\rangle^{\wp_o}_\ka\ .
\end{aligned}
\end{equation}
Finally we write $\de\wY=\smallint\de\wp_o\tfrac{\de}{\de\wp_o}\wY$ and  again integrate out~$\wp_i$ to define a flow operator
\begin{equation} \label{flowtilde}
\wR[\wp_o] \= \wR^{\textrm{inv}}[\wp_o] + \wR^{\textrm{gf}}[\wp_o] \=
-\tfrac{2\im}{\ka^2}\,\bigl\langle\,\smallint \wDelt_\al\ \smallint \wde_\al\wp_o\ \tfrac{\delta}{\delta\wp_o} \,\bigr\rangle
-\tfrac{2\im}{\ka^2}\,\bigl\langle\,\smallint \wDelt_{\textrm{GF}}\ \smallint \ws\wp_o\ \tfrac{\delta}{\delta\wp_o} \,\bigr\rangle
\end{equation}
and a multiplicative term
\begin{equation} \label{multtilde}
\wZ[\wp_o] \= \tfrac{2}{\ka^4}\,\bigl\langle\,\smallint \wDelt_\al\ \smallint \wde_\al\wLagr_{\textrm{GF}} \,\bigr\rangle
\end{equation}
as inner-field averages~$\langle\ldots\rangle$ at a fixed $\ka$~value.
In the transition from the third to the fourth line the supersymmetry and BRST Ward identities have been employed 
to move the supersymmetry and Slavnov variations to the functional~$\wY$. In the process, however, the supersymmetry variation
of the gauge-fixed action brings down the last term in the fourth line, which just multiplies~$\wY$, 
while the first three terms act as derivatives on~$\wY$.
Hence, at this stage our functional flow relation reads
\begin{equation} \label{tildeflow}
\ka\,\tfrac{\diff}{\diff \ka}\,\bigl\langle\,\wY[\wp_o]\,\bigr\rangle^{\wp_o}_\ka \=
\bigl\langle\,\bigl(\ka\,\tfrac{\pa}{\pa \ka}\ +\ \wR[\wp_o]\ +\ \wZ[\wp_o] \bigr)\,\wY[\wp_o]\,\bigr\rangle^{\wp_o}_\ka
\end{equation}
in the effective theory. If supersymmetry was global and thus independent of BRST~invariance, then
the multiplicative term could be recast into a third contribution to the flow operator,
\begin{equation} \label{Rmixtilde}
\wZ[\wp_o] \= \tfrac{2}{\ka^4}\,\bigl\langle\,\smallint \wDelt_\al\ \smallint \wde_\al\wLagr_{\textrm{GF}} \,\bigr\rangle
\= \tfrac{2}{\ka^4}\,\bigl\langle\,\smallint \wDelt_\al\ \smallint \wde_\al\wDelt_{\textrm{GF}}\ 
\smallint \ws\wp_o\ \tfrac{\delta}{\delta\wp_o}\,\bigr\rangle
\ =:\ \wR^{\textrm{mix}}[\wp_o]\ .
\end{equation}
However, due to the obstruction~\eqref{obstruction} two different multiplicative terms are created in the process,
and therefore we stay with~$\wZ$ here.

Since we are interested in a Nicolai map for Poincar\'e supergravity, we shall keep as outer fields only the
vierbein~$e^a_{\ \mu}$ and average over all other fields, including the conformal compensator~$\wP$.\footnote{
For a Nicolai map of conformal supergravity, $\wP$ should be retained as an outer field as well.}
In this way, \eqref{Mtilde} and~\eqref{susy} provide the gauge-invariant part of the flow operator,
\begin{equation}
\wR^{\textrm{inv}}[e^{\cdot}_{\ \cdot}] 
\= -\tfrac{2\im}{\ka^2} \bigl\langle\,\smallint \wDelt_\al\ \smallint \wde_\al e^a_{\ \mu}\,\bigr\rangle
\,\tfrac{\de}{\de e^a_{\ \mu}}
\= \tfrac{\im}{\ka^2} \bigl\langle\,
\smallint e\,\textrm{Re}\,\bigl[\tfrac{1}{\sqrt{2}}\wP\,P_L\widetilde{\slashed\nabla}\wro\bigr]_\al\ 
\smallint \bigl[ \bar\wps_\mu\ga^a\bigr]_\al\,\bigr\rangle\,\tfrac{\de}{\de e^a_{\ \mu}} \ ,
\end{equation}
where we used the auxiliary-field equation of motion~$\wF=0$. 
Likewise, the U(1) gauge field vanishes on-shell, $\widetilde{A}_\mu=0$,
and the dilatation gauge field will be gauge-fixed to zero, $\widetilde{b}_\mu=0$.\footnote{
This is not quite correct because, with gauge fixing, the auxiliary equations of motion yield
$\wF\sim\widetilde{c}^* P_L\widetilde{c}\ $ and $\ \widetilde{A}_\mu\sim\widetilde{c}^*\ga_\mu\widetilde{c}\ $ 
with the spinor ghost~$\widetilde{c}\equiv\widetilde{c}^Q$. As a consequence, quartic ghost terms appear in the action,
which however contribute only at $O(\hbar^2)$ to the Nicolai map. We ignore this ``quantum dressing'' for the time being,
but it can be handled along the lines of~\cite{CLR}.}

At this stage we may pass to the perturbative fields (without tilde), obtained from tilded fields by a rescaling,
\begin{equation}
e^a_{\ \mu} = \de_\mu^a + \ka\,\vp^a_{\ \mu}\ ,\quad
\wps_\mu^\al = \ka\,\psi_\mu^\al\ ,\quad
\wP = 1 + \ka\,\phi\ ,\quad
\wro^\al = \ka\,\rho^\al\ ,\quad
\wF = \ka\,F\ ,
\end{equation}
and likewise for the other connections,
with a nonzero background for the frame field and the scalar compensator,
so that all bosons, fermions and auxiliaries have a canonical mass dimension $1$, $\tfrac32$ and~$2$, respectively.
We note that our $\vp^a_{\ \mu}$ denotes the frame deviation from Minkowski space,
and $\phi$ describes the deviation from the Poincar\'e supergravity value.
In the process we observe that
\begin{equation}
\Lagr[\phi_o,\phi_i]\ :=\ \tfrac1{\ka^2}\wLagr[\wp_o,\wp_i] \quad\und\quad  \Delt[\phi_o,\phi_i]\ :=\ \tfrac1{\ka^2}\wDelt[\wp_o,\wp_i]
\end{equation}
are well defined for $\ka{\to}0$, so that the explicit inverse $\ka$~powers in \eqref{flowtilde} and~\eqref{multtilde} are absorbed. 
A potential $O(\tfrac1\ka)$ term is proportional to $\slashed\pa\rho$ and hence absent as a total derivative under the integral.
Paying attention to the identity
\begin{equation}
\ka\,\tfrac{\pa}{\pa\ka}\,\wY[\wp_o]\big|_{\wp_o=v+\ka\phi_o} \=
\ka\,\tfrac{\pa}{\pa\ka}\,Y[\phi_0] - \smallint\phi_o\,\tfrac{\de}{\de\phi_o}\,Y[\phi_0] \=
\bigl( \ka\,\tfrac{\pa}{\pa\ka} - E \bigr)\,Y[\phi_0]
\end{equation}
with the Euler operator~$E$ and a background value~$v$, our flow equation~\eqref{tildeflow} becomes
\begin{equation} \label{pertflow}
\ka\,\tfrac{\diff}{\diff \ka}\,\bigl\langle\,Y[\phi_o]\,\bigr\rangle^{\phi_o}_\ka \=
\bigl\langle\,\bigl(\ka\,\tfrac{\pa}{\pa \ka}\,-\,E[\phi_o]\,+\,R^{\textrm{inv}}[\phi_o]\,+\,R^{\textrm{gf}}[\phi_o]\,+\,Z[\phi_o] \bigr)\,
Y[\phi_o]\,\bigr\rangle^{\phi_o}_\ka\ .
\end{equation}
Inserting the supergravity expressions for $\phi_o=\vp^\cdot_{\ \cdot}$ and $v=\de_\cdot^\cdot$, the ingredients are
\begin{equation} \label{RZpert}
\begin{aligned}
R^{\textrm{inv}} &\= \im\,\bigl\langle\,\smallint \Delt_\al\ \smallint [ \bar\psi_\mu\ga^a ]_\al \,\bigr\rangle\,\tfrac{\de}{\de \vp^a_{\ \mu}}
\qquad\qquad\qquad\qquad \=  R_0^{\textrm{inv}} + \ka\,R_1^{\textrm{inv}} + \ka^2 R_2^{\textrm{inv}} + \ldots\ ,\\
R^{\textrm{gf}} &\= \im\,\bigl\langle\,\smallint c^{*A}\,F_A\ \smallint s\,\vp^a_{\ \mu}\,\bigr\rangle\,\tfrac{\de}{\de \vp^a_{\ \mu}}
\qquad\qquad\qquad\qquad \=  R_0^{\textrm{gf}} + \ka\,R_1^{\textrm{gf}} + \ka^2 R_2^{\textrm{gf}} + \ldots\ ,\\
Z &\= 2\,\bigl\langle\,\smallint\Delt_\al\ \smallint\bigl( -\ve_A F_A T^A\,\de_\al F_A - c^{*A}\de_\al s F_A\bigr)\,\bigr\rangle
\= Z_0 + \ka\,Z_1 + \ka^2 Z_2 + \ldots\ ,
\end{aligned}
\end{equation}
with $\ve_A{=}+1$ or $-1$ for bosonic and fermionic symmetries, respectively, 
\begin{equation} \label{Mpert}
{\Delt_\al}\big|_{F=0} \= \tfrac1{\sqrt{2}}\,\ep^{\tr\ln(\de_{\cdot}^{\cdot}+\ka\vp^{\cdot}_{\ \cdot})}\,
\textrm{Re}\bigl[(\tfrac1\ka{+}\phi)\,P_L\,\slashed\nabla\rho\bigr]_\al
\end{equation}
and~\footnote{
The $O(\ka^0)$ part was missed in~\cite{AKL}.}
\begin{equation} \label{spert}
s\,\vp_{a\mu} \= \de_a^\nu\pa_\mu c^P_\nu - \eta_{a\mu} c^D +
\ka\,(c^P_\nu\pa^\nu\vp_{a\mu} + \vp_a^{\ \nu}\pa_\mu c^P_\nu - c^M_{ab}\vp^b_{\ \mu} - c^D \vp_{a\mu} - \tfrac12\bar\psi_\mu\ga_a c^Q)\ .
\end{equation}
Implicit $\ka$ dependence hides in $\slashed\nabla$ and in the inner-field correlators~$\langle\ldots\rangle$.
The bracket here refers to functional averaging over $\phi$, $\rho$, $\psi$ and all auxiliary and ghost fields.
It produces propagators for the $(\phi,\rho,\psi)$ system in the vierbein background but also four-point functions
due to the (hidden) fermion-cubed terms in~$\slashed\nabla\rho$, which will however not contribute to the classical part of the Nicolai map.
Having eliminated~$F$ and Nakanishi--Lautrup auxiliaries via $b^A=T^AF_A$, the Lagrangian governing the inner bracket is
\begin{equation} \label{Lsusygf}
\Lagr_{\textrm{SUSY+GF}} \=
e(\vp^{\cdot}_{\ \cdot})\,\textrm{Re}\bigl[
(\tfrac1\ka{+}\phi)\nabla_a\nabla^a(\tfrac1\ka{+}\phi) - \bar\rho P_L\slashed\nabla\rho +
\tfrac{1}{\sqrt{2}}(1{+}\ka\phi)\bar{\slashed\psi} P_L\slashed\nabla\rho \bigr]
- \tfrac12 F_A\,T^A F_A - (s\,F_A)\,c^{*A}\ .
\end{equation}
This action is regular at $\ka{=}0$ and can be Taylor expanded.
The perturbative flow equation~\eqref{pertflow} is not obviously regular at~$\ka{=}0$,
although it must be so on general grounds. Dividing by~$\ka$ and expanding, it reads
\begin{equation} \label{singularflow}
\tfrac{\diff}{\diff \ka}\,\bigl\langle\,Y\,\bigr\rangle^{\phi_o}_\ka \=
\bigl\langle\,\tfrac{\pa}{\pa \ka}\,Y\ +\ \tfrac1\ka\bigl( R_0^{\textrm{inv}}\,+\,R_0^{\textrm{gf}}\,-\,E\,+\,Z_0 \bigr)\,Y
\ +\ \bigl( R_1^{\textrm{inv}}\,+\,R_1^{\textrm{gf}}\,+\,Z_1 \bigr)\,Y\ +\ O(\ka)\,\bigr\rangle^{\phi_o}_\ka\ ,
\end{equation}
which implies that
\begin{equation} \label{consistency}
\bigl\langle\,\bigl( R_0^{\textrm{inv}}\,+\,R_0^{\textrm{gf}}\,-\,E\,+\,Z_0 \bigr)\,Y\,\bigr\rangle^{\phi_o}_{\ka=0} \= 0
\end{equation}
must be fulfilled in the free theory for consistency.

\section{What is the matter with multiplicative terms?}\label{sec:multiplicative}

At first sight, it seems that the multiplicative $Z$ contribution in~\eqref{pertflow} is detrimental to the regularity 
condition~\eqref{consistency} because it is not compatible with a functional solution,
\begin{equation}
R_0^{\textrm{inv}}\,+\,R_0^{\textrm{gf}}\,-\,E\,+\,Z_0 \ \buildrel{?}\over{=}\ 0\ .
\end{equation}
However, inside the final free correlator $\langle\ldots\rangle_{\ka=0}^{\phi_o}$ such a term can potentially be rewritten 
as a functional derivative, as we shall demonstrate now.
How this is achieved we can learn from the simpler case of supersymmetric Yang--Mills theory~\cite{ALMNPP},
consisting of a Lie-algebra valued gauge potential~$A_\mu$, a Majorana gaugino~$\la$ and a real auxiliary~$D$, 
in an axial gauge $F[A]=n{\cdot}A=0$ with a fixed four-vector~$n^\mu$.
Perturbative and non-perturbative field scalings are related via $\wA_\mu=g\,A_\mu$, $\wla=g\,\la$, $\widetilde{D}=g\,D$,
and likewise for the ghost, antighost and Nakanishi--Lautrup auxiliary $c$, $\bar{c}$ and~$b$, respectively, 
with a dimensionless gauge coupling~$g$.
The off-shell gauge-fixed action is a combination of a supervariation and a Slavnov variation,
\begin{equation}
\wLagr_{\textrm{SUSY}} + \wLagr_{\textrm{GF}} \=  \wde_\al\,\wDelt_\al + \ws\,\wDelt_{\textrm{GF}} 
\end{equation}
with
\begin{equation}
\wDelt_\al \= \tfrac18\bigl\{ -\tfrac12 \wF_{\mu\nu}\ga^{\mu\nu}\wla + \widetilde{D}\,\ga_5\wla \bigr\}_\al \und
\wDelt_{\textrm{GF}} \= \bar{\widetilde{c}}\,\bigl(  \tfrac1{2\xi} \widetilde{b} - n{\cdot}\wA \bigr) \ ,
\end{equation}
where the color trace is implicit and a gauge parameter~$\xi$ specifies the quantum gauge.
At this stage we may pass to the perturbative fields and drop the tildes.
Because $\Lagr_{\textrm{GF}}$ is not supersymmetric,
\begin{equation}
\de_\al \Lagr_{\textrm{GF}} \= 
-b\;n^\mu\de_\al A_\mu - \bar{c}\;n^\mu\de_\al D_\mu c \=
\bigl(\xi\,n{\cdot} A+g\,\bar{c}\ c\bigr)\,(\bar\la\,n{\cdot}\ga)_\al
\end{equation}
using $b=\xi\,n{\cdot}A$, $sA_\mu=D_\mu c=\pa_\mu c+g\,A_\mu c$, $\de_\al A_\mu=-(\bar\la\ga_\mu)_\al$
and implicit color factors, also here the flow equation features a multiplicative term,
\begin{equation}
Z \= 2\,\bigl\langle\,\smallint \Delt_\al\ \smallint \de_\al \Lagr_{\textrm{GF}} \,\bigr\rangle \=
-\tfrac18\,\bigl\langle\,\smallint F_{\mu\nu}(\ga^{\mu\nu}\la)_\al \
\smallint \bigl(\xi\;n{\cdot}A+g\,\bar{c}\ c\bigr)\,(\bar\la\,n{\cdot}\ga)_\al \,\bigr\rangle\ ,
\end{equation}
where we have integrated out the auxiliary via~$D{=}0$, and the bracket still averages over
the gaugino and the ghost-antighost system.

Because here supersymmetry is global, it commutes with any gauge transformation, and hence
\begin{equation}
\de_\al\,s = - s\,\de_\al \und s\,\Delt_\al = 0 \ .
\end{equation}
As already mentioned in~\eqref{Rmixtilde}, this allows us to recast the multiplicative term above via
\begin{equation}
\bigl\langle\bigl\langle\,\smallint \Delt_\al\ \smallint\de_\al\,\Lagr_{\textrm{GF}}\ \ Y \,\bigr\rangle\bigr\rangle^A_g \=
-\bigl\langle\bigl\langle\,\smallint \Delt_\al\ \smallint s\;\de_\al\,\Delt_{\textrm{GF}}\ \ Y \,\bigr\rangle\bigr\rangle^A_g \=
\bigl\langle\bigl\langle\,\smallint \Delt_\al\ \smallint \de_\al\,\Delt_{\textrm{GF}}\ \ s\,Y \,\bigr\rangle\bigr\rangle^A_g\ ,
\end{equation}
which again acts as a derivation on~$Y$ and thus produces merely a second extension to the flow operator,
\begin{equation}
R \= R^{\textrm{inv}} + R^{\textrm{gf}} + R^{\textrm{mix}} \qquad\with\quad
R^{\textrm{mix}}[A] \=
2\,\bigl\langle\,\smallint \Delt_\al\ \smallint \de_\al\,\Delt_{\textrm{GF}}\ 
\smallint s\,A_\mu\ \tfrac{\delta}{\delta A_\mu}\,\bigr\rangle\ .
\end{equation}
It is satisfying to check that this total flow annihilates $n{\cdot}A$, because
$R^{\textrm{gf}}+R^{\textrm{mix}}$ projects $Y$ onto the gauge slice~$F{=}0$.

Comparing the two versions, it follows from 
$\langle Z\ Y\rangle_g^A=\langle R^{\textrm{mix}}\,Y\rangle_g^A$ that
\begin{equation} \label{multder}
\bigl\langle\, \smallint F_{\mu\nu}(\ga^{\mu\nu}\la)_\al \ \smallint 
\bigl(\xi\;n{\cdot}A+g\,\bar{c}\ c\bigr)\,(\bar\la\,n{\cdot}\ga)_\al\ Y\,\bigr\rangle_g^A \=
\bigl\langle\,\smallint F_{\mu\nu}(\ga^{\mu\nu}\la)_\al\ \smallint (\bar\la\,n{\cdot}\ga)_\al\,\bar{c}\ 
\smallint D_\mu c\,\tfrac{\de Y}{\de A_\mu} \,\bigr\rangle_g^A\ ,
\end{equation}
which equates a multiplicative with a derivative term for the final path integral over~$A$,
but not as an identity for the effective theory inside the bracket.
The inverse Nicolai map is ultimately applied in the free theory, $g{=}0$, where the relation~\eqref{multder}
becomes
\begin{equation}
\bigl\langle\, \smallint \pa_\mu A_\nu(\ga^{\mu\nu}\la)_\al \ \smallint 
(\bar\la\,n{\cdot}\ga)_\al\,\xi\;n{\cdot}A\ \ Y\,\bigr\rangle_0^A \=
\bigl\langle\,\smallint \pa_\mu A_\nu(\ga^{\mu\nu}\la)_\al\ \smallint (\bar\la\,n{\cdot}\ga)_\al\,\bar{c}\ 
\smallint \pa_\mu c\;\tfrac{\de Y}{\de A_\mu} \,\bigr\rangle_0^A
\end{equation}
Employing the free propagators
\begin{equation}
\bcontraction{}{\la}{\quad}{\bar\la}\la\quad\bar\la \= \im\,\slashed\pa\Box^{-1}  \und
\bcontraction{}{\bar{c}}{\quad}{c}\bar{c}\quad c \= \im\,(n{\cdot}\pa)^{-1}
\end{equation}
and partially integrating $\pa_\mu$, we arrive at a relation in the effective (free) gauge theory,
\begin{equation}
\bigl\langle\,\smallint\!\smallint \pa_\mu A_\nu\,\tr(\ga^{\mu\nu}\,\slashed\pa\Box^{-1} n{\cdot}\ga)\,\xi\;n{\cdot}A\ \ Y\,\bigr\rangle_0^A \=
-\im\,\bigl\langle\,\smallint\!\smallint\!\smallint \pa_\mu A_\nu\,\tr(\ga^{\mu\nu}\,\slashed\pa\Box^{-1} n{\cdot}\ga)\,(n{\cdot}\pa)^{-1}
\pa_\mu\tfrac{\de Y}{\de A_\mu} \,\bigr\rangle_0^A\ .
\end{equation}
Performing the gamma traces $\tr(\ga^{\mu\nu}\slashed\pa\,n{\cdot}\ga)=4(n^\mu\pa^\nu{-}n^\nu\pa^\mu)$ one finds that
\begin{equation} \label{finalfree}
\bigl\langle\,\smallint n{\cdot}A^{\perp}\ \xi\,n{\cdot}A \ \ Y\,\bigr\rangle_0^A \=
-\im\,\bigl\langle\,\smallint\!\smallint n{\cdot}A^{\perp}\,(n{\cdot}\pa)^{-1} \pa_\mu\tfrac{\de Y}{\de A_\mu} \,\bigr\rangle_0^A
\end{equation}
with the transverse gauge potential $A^\perp=A-\pa\Box^{-1}\pa{\cdot}A$.

The evaluation of these free-field correlators invokes Wick's theorem.
On the left-hand side, $\xi\,n{\cdot}A$ is to be contracted once with $n{\cdot}A^\perp$ and once with~$Y$ inside the bracket.
However, the axial-gauge propagator~\cite{IZ}
\begin{equation}
\xi\;n{\cdot}\bcontraction{}{A}{\quad}{A}A\quad A_\mu \= -\im\,(n{\cdot}\pa)^{-1}\pa_\mu
\end{equation}
implies that $\xi\;n{\cdot}\bcontraction{}{A}{\quad}{A}A\quad A^\perp=0$, and thus the left correlator in~\eqref{finalfree}
reduces to
\begin{equation}
\bigl\langle\,\smallint n{\cdot}A^{\perp}\ \xi\,n{\cdot}A \ \ Y\,\bigr\rangle_0^A \=
\bigl\langle\,\smallint n{\cdot}A^{\perp}\ \xi\,n{\cdot}\bcontraction{}{A}{\ \smallint}{A}A\ \smallint A_\mu\tfrac{\de Y}{\de A_\mu}
\,\bigr\rangle_0^A \=
-\im\,\bigl\langle\,\smallint\!\smallint n{\cdot}A^{\perp}\ (n{\cdot}\pa)^{-1} \pa_\mu \tfrac{\de Y}{\de A_\mu} \,\bigr\rangle_0^A 
\end{equation}
which indeed agrees with the right correlator! The technical reason for the match is an identity between the free
gluon and ghost propagator,
\begin{equation}
\xi\;n{\cdot}\bcontraction{}{A}{\quad}{A}A\quad A_\mu \= \bcontraction{}{\bar{c}}{\quad \pa_\mu}{c}\bar{c}\quad \pa_\mu c \ .
\end{equation}
Since it is known~\cite{LR1,MN,LR2} for supersymmetric Yang--Mills theory that
\begin{equation}
\bigl\langle\, \bigl( R_0^{\textrm{inv}}\,+\,R_0^{\textrm{gf}}\,+\,R_0^{\textrm{mix}}\,-\,E \bigr)\ Y \,\bigr\rangle_{g=0}^A \= 0\ ,
\end{equation}
this identity is at the same time instrumental for the regularity of the perturbation theory around~$g{=}0$.
Moreover, since it holds for any functional~$Y$, we can ignore the $\tfrac1{\ka}$ term
in~\eqref{singularflow} altogether and begin the perturbative expansion of $R$ and~$Z$ at order~$\ka$,
\begin{equation}
\tfrac{\diff}{\diff \ka}\,\bigl\langle\,Y\,\bigr\rangle^{\phi_o}_\ka \=
\bigl\langle\,\tfrac{\pa}{\pa \ka}\,Y\ +\ {\textstyle\sum}_{n=1}^{\infty} \ka^{n-1}
\bigl( R_n^{\textrm{inv}}\,+\,R_n^{\textrm{gf}}\,+\,Z_n \bigr)\,Y\,\bigr\rangle^{\phi_o}_\ka\ .
\end{equation}

The lesson of this example is that multiplicative terms in the flow equation might be converted into derivative ones.
The key insight is not to look for functional identities in the effective theory but to rewrite its final free correlators
with the help of Wick's theorem.
We have shown how this works to all orders for super Yang--Mills and will see it below to leading order for supergravity.
Yet, to establish the all-order existence of a Nicolai map for Poincar\'e supergravity we need to find a universal
conversion mechanism for getting rid of all multiplicative terms in the flow equation.
We repeat that this issue is only a gauge-fixing one and does not affect the gauge-invariant part of the theory,
so perhaps it can also be overcome by a more clever way of handling the gauge redundancy in the path integral.

\section{Perturbative expansion}\label{sec:pertexp}

The flow operator and multiplicative terms are to be constructed in a power series in the coupling~$\ka$.
Besides the explicit $\ka$~dependence in \eqref{RZpert}, the chiral boson, fermion and ghost correlators 
from the brackets have to be expanded in~$\ka$, using the Feynman rules from the gauge-fixed Lagrangian~\eqref{Lsusygf}.
The gauge-fixed free propagators require inverting the bosonic, fermion and ghost bilinear forms in~$\Lagr_{\textrm{SUSY+GF}}$.
Since Re$\phi$ mixes with the vierbein trace $\vp\equiv\vp^a_{\ a}$ already at the free level, while Im$\phi$ does not,
it is convenient to split 
\begin{equation}
\phi \= \tfrac1{\sqrt{2}}\,(\phi_1+\im\,\phi_2)
\end{equation}
into two real scalars $\phi_1$ and~$\phi_2$.

Recalling that, on the linear level for the graviton, the harmonic gauge in~\eqref{gauges} simplifies to
\begin{equation} \label{harmonic}
F_P^\nu \= \pa^\nu\vp - 2\,\pa_\mu\vp^{\mu\nu} \ ,
\end{equation}
and the quadratic part of the gauge-fixed Lagrangian reads (remember we gauge-fixed $b_\mu{=}0$)
\begin{equation} \label{Ltotal0}
\begin{aligned}
\Lagr_{\textrm{SUSY+GF}}|_{\ka=0}
&\= \tfrac16 \vp^{ab}\Box\vp_{ab} - \tfrac13\vp^{ab}\pa_b\pa^c\vp_{ac} + \tfrac13\vp^{ab}\pa_b\pa_a\vp - \tfrac16\vp\Box\vp 
- \tfrac1{12}\xi_P(\pa_a\vp{-}2\pa^b\vp_{ba})^2 \\
&\quad -\ \tfrac12\phi_1\Box\phi_1  - \tfrac12\phi_2\Box\phi_2 - \tfrac1{3\sqrt{2}}\phi_1\Box\vp 
- \tfrac12\xi_D\phi_1\Box\phi_1 - \tfrac12\xi_Y\phi_2\Box\phi_2 \\
&\quad -\ \tfrac16\bar\psi_a\ga^{abc}\pa_b\psi_c +\tfrac12\bar\rho\slashed\pa\rho -\tfrac{\sqrt{2}}{3}\bar\psi_a\ga^{ab}\pa_b\rho
+\tfrac16\xi_Q\bar{\slashed\psi}\slashed\pa\slashed\psi +\tfrac12\xi_S\bar\rho\slashed\pa\rho \\[2pt] 
&\quad +\ c_P^{*\mu}\Box c^P_\mu + c_M^{*ab}c^M_{ab} + c_M^{*ab}\de_a^\mu\pa_b c^P_\mu
+ c_K^{*\mu}c^K_\mu + \tfrac12 c_K^{*\mu}\pa_\mu c^D + c_D^* c^D + c_D^*\pa^\mu c^P_\mu + c_Y^* c^Y \\
&\quad +\ \bar{c}_Q\slashed\pa c^Q + \bar{c}_S c^S - 4\,\bar{c}_Q c^S + A^a A_a - \bar{F}\,F \\[4pt]
&\= \tfrac16 \vp^{ab}\Box\vp_{ab}-\tfrac{1}{12}\vp\Box\vp - \tfrac{1}{12}(\xi_P{-}1)(\pa_a\vp{-}2\pa^b\vp_{ba})^2 \\
&\quad -\ \tfrac12(1{+}\xi_D)\phi_1\Box\phi_1
- \tfrac12(1{+}\xi_Y)\phi_2\Box\phi_2 - \tfrac1{3\sqrt{2}}\phi_1\Box\vp \\
&\quad +\ \tfrac{1}{12}\bar\psi_a\ga^c\slashed\pa\ga^a\psi_c - \tfrac{1}{12}(1{-}2\xi_Q)\bar{\slashed\psi}\slashed\pa\slashed\psi
+\tfrac12(1{+}\xi_S)\bar\rho\slashed\pa\rho -\tfrac{\sqrt{2}}{3}\bar\psi_a\ga^{ab}\pa_b\rho \\[2pt]
&\quad +\ c_P^{*\mu}\Box c^P_\mu + c_M^{*ab}(c^M_{ab}-\pa_{[a}c^P_{b]}) 
+ c_K^{*\mu}(c^K_\mu+\tfrac12\pa_\mu c^D) + c_D^*(c^D+\pa^\mu c^P_\mu) + c_Y^* c^Y  \\
&\quad +\ \bar{c}_Q(\slashed\pa c^Q-4\,c^S) + \bar{c}_S c^S + A^a A_a - \bar{F}\,F \ ,
\end{aligned}
\end{equation}
suggesting a redefinition of the ghost (but not the antighost) fields.
Inverting this bilinear form yields the free-fermion propagators
\begin{equation} \label{fermiprop}
\begin{aligned}
\bcontraction{}{\rho}{\quad}{\bar\rho}\rho\quad\bar\rho &\= \tfrac{\im}{\xi_S}\,\slashed\pa\,\Box^{-1}\ ,\\
\bcontraction{}{\psi}{_c\quad}{\bar\rho}\psi_c\quad\bar\rho &\= -\tfrac{\im}{\xi_S\sqrt{2}}(\ga_c\slashed\pa-4\pa_c)\,\Box^{-1}\, \\
\bcontraction{}{\rho}{\quad}{\bar\psi}\rho\quad\bar\psi_d &\= \tfrac{\im}{\xi_S\sqrt{2}}(\slashed\pa\ga_d-4\pa_d)\,\Box^{-1}\ ,\\
\bcontraction{}{\psi}{_c\quad}{\bar\psi}\psi_c\quad\bar\psi_d &\= \im\,\bigl[ (\tfrac32{+}\tfrac{1}{2\xi_S})\ga_d\slashed\pa\ga_c
+(\tfrac{3}{\xi_Q}{-}\tfrac{8}{\xi_S}{-}6)\tfrac{\pa_c\pa_d}{\Box}\slashed\pa
+\tfrac{1}{\xi_S}(\eta_{cd}\slashed\pa+\ga_c\pa_d+\ga_d\pa_c)\bigr]\,\Box^{-1}\ .
\end{aligned}
\end{equation}
We remark that $\gamma{\cdot}\bcontraction{}{\psi}{\ }{\bar\rho}\psi\ \bar\rho
=\bcontraction{}{\rho}{\ }{\bar\psi}\rho\ \bar\psi{\cdot}\gamma=0$.
Obviously, a preferred choice is $\xi_Q=\tfrac12$ and $\xi_S\to\infty$,
which leaves only $\bcontraction{}{\psi}{_c\ }{\bar\psi}\psi_c\ \bar\psi_d=\tfrac{3\im}{2}\ga_d\tfrac{\slashed\pa}{\Box}\ga_c$.
For the bosons, in the limit $\xi_{D,Y}\to\infty$, only the free graviton propagator remains,
\begin{equation} \label{boseprop}
\bcontraction{}{\vp}{_{ab}\quad}{\vp}\vp_{ab}\quad\vp_{cd} \=
\tfrac{3\im}{2}\bigl[\eta_{ac}\eta_{bd}+\eta_{ad}\eta_{bc}-\eta_{ab}\eta_{cd} \bigr]\Box^{-1}
- \tfrac{3\im}{2}(\tfrac1{\xi_P}{-}1)\bigl[
\eta_{ac}\pa_b\pa_d+\eta_{ad}\pa_b\pa_c+\eta_{bc}\pa_a\pa_d+\eta_{bd}\pa_a\pa_c\bigr]\Box^{-2}
\end{equation}
Finally, the nonzero ghost propagators come out as
\begin{equation} \label{ghostprop}
\begin{aligned}
&\bcontraction{}{c}{^P_\mu\quad}{c}c^P_\mu\quad c_P^{*\nu} \= \im\de_\mu^\nu\,\Box^{-1}\ ,\qquad
\bcontraction{}{c}{^D\quad}{c}c^D\quad c_P^{*\nu} \= -\im\,\pa^\nu\Box^{-1}\ ,\qquad
\bcontraction{}{c}{^K_\mu\quad}{c}c^K_\mu\quad c_P^{*\nu} \= \im\,\pa_\mu\pa^\nu\,\Box^{-1}\ ,\\
&\bcontraction{}{c}{^M_{ab}\quad}{c}c^M_{ab}\quad c_M^{*cd} \= \tfrac{\im}{2}(\de_a^c\de_b^d-\de_a^d\de_b^c)\ ,\qquad
\bcontraction{}{c}{^M_{ab}\quad}{c}c^M_{ab}\quad c_P^{*\nu} \= -\tfrac{\im}{2}(\de_a^\nu\pa_b-\de_b^\nu\pa_a)\,\Box^{-1}\ ,\\
&\bcontraction{}{c}{^K_\mu\quad}{c}c^K_\mu\quad c_K^{*\nu} \= \im\,\de_\mu^\nu\ ,\qquad
\bcontraction{}{c}{^K_\mu\quad}{c}c^K_\mu\quad c_D^* \= -\tfrac{\im}{2}\,\pa_\mu\ ,\qquad
\bcontraction{}{c}{^D\quad}{c}c^D\quad c_D^* \= \im\ ,\qquad
\bcontraction{}{c}{^Y\quad}{c}c^Y\quad c_Y^* \= \im\ ,\\
&\bcontraction{}{c}{^Q\quad}{c}c^Q\quad \bar{c}_Q \= -\im\,\slashed\pa\,\Box^{-1}\ ,\qquad
\bcontraction{}{c}{^S\quad}{c}c^Q\quad \bar{c}_S \= -4\im\,\slashed\pa\,\Box^{-1}\ ,\qquad
\bcontraction{}{c}{^S\quad}{c}c^S\quad \bar{c}_S \= -\im\ .
\end{aligned}
\end{equation}
In order to set up a perturbative expansion of the flow equation, one must develop a power series for all full propagators
in the vierbein background, except for the vierbein itself. To this end, all interaction vertices involving inner fields~$\phi_i$
have to be collected from the superconformal Lagrangian. Furthermore, $\Delt_\al$ from~\eqref{Mpert} must be expanded in~$\ka$ as well.
With these ingredients we are in principle able to dive into a perturbative expansion of our candidate inverse Nicolai map.
Here, we just provide the leading order of the ingredient for~$R^{\textrm{inv}}$,
\begin{equation}
\begin{aligned}
\Delt &\= \smallint\,\bigl\{1{+}\ka\vp{+}O(\ka^2)\bigr\}\,\textrm{Re}\bigl\{ 
\tfrac1{\sqrt{2}} P_L \bigl(\bar{F} + (\tfrac1\ka{+}\phi)\,\slashed\nabla\bigr)\rho\ +\ 
\tfrac12\,(\tfrac1\ka{+}\phi)\,\bar{F}\ga^a P_R\psi_a \bigr\} \ + O(\ka) \\[4pt]
&\= P_L\Bigl[ \tfrac1{\sqrt{2}}\bigl(\phi\ga^a+\tfrac12\vp\ga^a-\tfrac12\vp^a_{\ b}\ga^b\bigr)\,\pa_a\rho +
\bigl(\tfrac16\vp+\tfrac13\phi-\tfrac16\bar\phi\bigr)\ga^\mu\ga^\nu\pa_\mu\psi_\nu +
\bigl(\tfrac23\bar\phi-\tfrac13\phi\bigr)\,\pa^\mu\psi_\mu \\
&\qquad -\
\vp_{(ab)}\,\bigl(\tfrac16\ga^a\ga^c\pa^b\psi_c+\tfrac1{36}\ga^c\ga^b\pa_c\psi^a-\tfrac1{36}\pa^a\psi^b\bigr) \\[2pt]
&\qquad +\
\vp_{[b\,c]}\,\bigl(\tfrac1{12}\ga^{\mu\nu bc}\pa_\mu\psi_\nu-\tfrac16\ga^\mu\ga^c\pa_\mu\psi^b+
\tfrac16\ga^\mu\ga^c\pa^b\psi_\mu-\tfrac12\pa^b\psi^c+\tfrac1{4\sqrt{2}}\ga^{\mu bc}\pa_\mu\rho\bigr) \\
&\qquad -\
\tfrac{\im}{2\sqrt{2}}\ga^\mu A_\mu\rho + \tfrac12\ga^\mu\ga^\nu(b_\nu{-}\im A_\nu)\psi_\mu - \tfrac13\ga^\mu\ga^\nu b_\nu\psi_\mu -
\tfrac16 b^\mu\psi_\mu + \tfrac{\im}{2}\ga^{\mu\nu}A_\mu\ga_5\psi_\nu + \tfrac1{\sqrt{2}}F\rho \Bigr] \ + O(\ka)\ ,
\end{aligned}
\end{equation}
where (anti)symmetrizations are done with weight~$\tfrac12$.
We have checked that the (traced) supervariation of the above expression indeed produces the full quadratic part~\eqref{Ltotal0} of
the superconformal Lagrangian without the gauge-fixing terms.
Putting $F$, $b_\mu$, $A_\mu$ and $\vp_{[b\,c]}$ to zero and imposing the harmonic gauge condition $F_P{=}0$ from~\eqref{harmonic}
in the Landau gauge~$\xi_P{\to}\infty$, this simplifies to
\begin{equation} \label{Mquad}
\begin{aligned}
\Delt\big|_{\textrm{Landau}} 
&\= \smallint\,\bigl\{1{+}\ka\vp{+}O(\ka^2)\bigr\}\,\textrm{Re}\bigl\{\tfrac{1}{\sqrt{2}}(\tfrac1\ka{+}\phi)\,P_L\slashed\nabla\rho\bigr\} 
\ + O(\ka) \\[4pt]
&\= \textrm{Re} \smallint P_L \bigl\{
\tfrac{1}{\sqrt{2}}(\phi\,\ga^a+\tfrac12\vp\,\ga^a-\tfrac12\vp^a_{\ b}\ga^b)\,\pa_a\rho 
+\tfrac13([\phi{-}\tfrac12\bar\phi]\ga^a\ga^b - [\phi{-}2\bar\phi]\eta^{ab})\,\pa_a\psi_b \\
&\qquad\qquad\quad +\,\tfrac16(\vp\,\ga^a\ga^b-\vp^{ab}-\vp^a_{\ c}\ga^c\ga^b+\vp^b_{\ c}\ga^c\ga^a)\,\pa_a\psi_b\bigr\}\ + O(\ka)\\[4pt]
&\= \smallint\,\bigl\{
\tfrac14 (\phi_1\ga^a+\im\ga_5\phi_2\ga^a+\tfrac1{\sqrt{2}}\vp\,\ga^a-\tfrac1{\sqrt{2}}\vp^a_{\ b}\ga^b)\,\pa_a\rho 
+\tfrac1{4\sqrt{2}}\,\im\ga_5\phi_2(\ga^a\ga^b{-}2\eta^{ab})\,\pa_a\psi_b \\
&\qquad\ +\,\tfrac{1}{12} (\tfrac1{\sqrt{2}}\phi_1\ga^a\ga^b+\sqrt{2}\phi_1\eta^{ab}
+\vp\,\ga^a\ga^b-\vp^{ab}-\vp^a_{\ c}\ga^c\ga^b+\vp^b_{\ c}\ga^c\ga^a)\,\pa_a\psi_b \bigr\} \ + O(\ka)\ .
\end{aligned}
\end{equation}
This is used below to generate the leading term~$R_0^{\textrm{inv}}$ of the flow operator.

\section{Consistency check}\label{sec:check}

As a first test of the method, let us check the regularity condition~\eqref{consistency}
in the Landau gauge $\xi_A{\to}\infty$ for~$A=P,D,Y,S$.\footnote{ 
Remember that we already took $\xi_{M,K}{\to}\infty$ but did not fix $\xi_Q$ yet.}
At this order, all composite connections as well as auxiliary fields vanish.
We need to take~\eqref{Mquad} and use the free propagators~\eqref{fermiprop} for $\xi_S{\to}\infty$ 
to evaluate the contraction with $\bar\psi_\mu$ in~\eqref{RZpert}.
The gamma traces remove all $\phi_2$ contributions.
The chiral boson average replaces $\phi_1$ and $\phi_2$ by their classical values
\begin{equation} \label{phicl}
\phi_1\ \to\ -\vp/3\sqrt{2}(1{+}\xi_D) \und \phi_2\ \to\ 0 \ .
\end{equation}
After a straightforward but lengthy computation one arrives at
\begin{align}
R_0^{\textrm{inv}} &\= \smallint\bigl\{ \vp_{ab} - \tfrac12\vp\,\eta_{ab} + (1{-}\tfrac1{\xi_Q})\vp\tfrac{\pa_a\pa_b}{\Box}
+(2{-}\tfrac1{\xi_Q})\vp_{ac}\tfrac{\pa^c\pa_b}{\Box} + \tfrac12\vp_{cd}\tfrac{\pa^c\pa^d}{\Box}\eta_{ab}
+(\tfrac2{\xi_Q}{-}4)\vp_{cd}\tfrac{\pa^c\pa^d\pa_a\pa_b}{\Box\quad\Box} \bigr\} \tfrac{\de}{\de\vp_{ab}} \nn \\[2pt]
&\quad +\,\tfrac1{1{+}\xi_D} \smallint\bigl\{ \tfrac16\vp\,\eta_{ab} + (\tfrac1{2\xi_Q}{-}\tfrac23)\,\vp\,\tfrac{\pa_a\pa_b}{\Box}
\bigr\} \tfrac{\de}{\de\vp_{ab}}\ .
\end{align}
In the Landau gauge $\xi_{P,D}{\to}\infty$, the second line vanishes and, with $F^\nu_P{=}0$, the first line simplifies to
\begin{equation} \label{R0inv}
R_0^{\textrm{inv}}\big|_{\textrm{Landau}} \= \smallint\,\bigl\{ \vp_{ab}
\,-\,\tfrac14\,\vp\,\eta_{ab}\,-\,\tfrac1{2\xi_Q}\,\vp\,\tfrac{\pa_a\pa_b}{\Box}\bigr\} \tfrac{\de}{\de\vp_{ab}}\ ,
\end{equation}
which indeed agrees with (the invariant part of) the $\ka{=}0$ flow operator for Poincar\'e supergravity~\cite{AKL}.

Next, we consider the leading part of $R^{\textrm{gf}}$,
\begin{equation}
R_0^{\textrm{gf}} \= \im\,\bigl\langle\,\smallint c^{*A}\,F_A\ 
\smallint \bigl[\de_a^\nu\pa_\mu c^P_\nu - \eta_{a\mu} c^D \bigr]\,\bigr\rangle\,\tfrac{\de}{\de \vp_{a\mu}}\ .
\end{equation}
From the ghost propagators~\eqref{ghostprop} we see that only $A=P,M,K,D$ contribute to the $\bcontraction{}{c}{^{*A}\ }{c}c^{*A}\ c^B$
contractions, but with the Landau gauge for $M,K,D$ we are left with
\begin{equation}
R_0^{\textrm{gf}}\big|_{\xi_{M,D,Y}\to\infty} \= -\im\,\bigl\langle\,\smallint (\pa^b\vp-2\pa^c\vp^b_{\ c})\,c_b^{*P}\ 
\smallint c_a^P \,\bigr\rangle\;\pa_\mu\tfrac{\de}{\de \vp_{a\mu}} \=
\smallint (\pa^a\vp-2\pa^c\vp^a_{\ c})\,\Box^{-1} \pa_b \tfrac{\de}{\de \vp_{ab}}\ .
\end{equation}
This remains proportional to the harmonic gauge condition, and thus
\begin{equation}
R_0^{\textrm{gf}}\big|_{\textrm{Landau}} \= 0\ .
\end{equation}

Finally, we have to take account of the $\ka{\to}0$ limit of multiplicative piece
\begin{equation}
Z\= 2\,\bigl\langle\,\smallint \Delt_\al\ \smallint\bigl( -\ve_A F_A\,T^A \,\de_\al F_A - c^{*A}\de_\al s F_A\bigr)\,\bigr\rangle\ .
\end{equation}
Because $s\,{F_A}|_{\ka=0}$ is pure ghost, and ghosts do not transform under supersymmetry, the ghost term does not contribute.
The other term vanishes on the gauge slice $F_A{=}0$, except for $A=D,Y,Q,S$ where $F_A$ gets contracted 
under the bracket. However, since $\de_\al (F_D{+}\im\,F_Y)\sim\bar\rho$ and free $\rho$ propagators~\eqref{fermiprop}
vanish for $\xi_S{\to}\infty$, only the fermionic gauge terms $A=Q,S$ survive,
\begin{equation}
\begin{aligned}
Z\big|_{F_A=0} &\= 
2\,\bigl\langle \smallint\,\Delt_\al\big|_{F=0}\,\smallint\,\bigl[ F_Q T^Q \de_\al F_Q + F_S T^S \de_\al F_S \bigr] \,\bigr\rangle \\[2pt]
&\= 2\,\bigl\langle \smallint\,\Delt_\al\big|_{F=0}\,\smallint\,\bigl[
\tfrac13\xi_Q \gamma{\cdot}\bar\psi\,\slashed\pa\,\de_\al\ga{\cdot}\psi+\xi_S\bar\rho\,\slashed{\pa}\,\de_\al\rho \bigr] \,\bigr\rangle \\[2pt]
&\= 2\,\bigl\langle \smallint\,\Delt_\al\big|_{F=0}\,\smallint\,\bigl[
\tfrac13\xi_Q \gamma{\cdot}\bar\psi\,\slashed\pa\,\tfrac14\gamma^\mu\om_{\mu ab}\ga^{ab} 
+\tfrac12\xi_S \bar\rho\,\slashed\pa\,\slashed\nabla\phi_1 \bigr] \,\bigr\rangle \ .
\end{aligned}
\end{equation}
With the help of
\begin{equation}
\ga^\mu\om_{\mu a b}\ga^{ab} = 2(\slashed\pa\vp-\ga^a\pa^b\vp_{ab}) +O(\ka) \und
\slashed\nabla\phi_1 = \slashed\pa\phi_1 + O(\ka)
\end{equation}
as well as the contractions
\begin{equation}
\tfrac13\xi_Q\,\bcontraction{}{\psi}{_b\quad}{\bar{\slashed\psi}}\psi_b\quad\bar{\slashed\psi}\ \to\ \im\,\pa_b\Box^{-1}\ ,\qquad
\xi_S\,\bcontraction{}{\rho}{\quad}{\bar\rho}\rho\quad\bar\rho\ \to\ \im\,\slashed\pa\Box^{-1}\ ,\qquad
\xi_S\,\bcontraction{}{\psi}{_b\quad}{\bar\rho}\psi_b\quad\bar\rho\ \to\ -\tfrac{\im}{\sqrt{2}}(\ga_b\slashed\pa-4\pa_a)\Box^{-1}
\end{equation}
in the limit $\xi_S\to\infty$ and borrowing $M\big|_{\textrm{Landau}}$ from \eqref{Mquad}, after a lengthy computation we get
\begin{equation}
Z_0\big|_{F_A=0}  \= \tfrac{\im}{3}\,\bigl\langle\,\smallint \bigl\{
\vp_{ab}\bigl(\eta^{ab}{-}\tfrac{\pa^a\pa^b}{\Box}\bigr)\,\Box\,\bigl(\eta^{cd}{-}\tfrac{\pa^c\pa^d}{\Box}\bigr)\,\vp_{cd}
+\tfrac{7}{\sqrt{2}}\,\phi_1\,\Box\bigl(\eta^{cd}{-}\tfrac{\pa^c\pa^d}{\Box}\bigr)\,\vp_{cd}
+6\,\phi_1\Box\phi_1 \bigr\}\,\bigr\rangle\ .
\end{equation}
Ultimately, we must average over $\phi_1$ and $\phi_2$, using the relevant contractions~\eqref{boseprop} and the 
equations of motion~\eqref{phicl}, but the Landau gauge $\xi_D{+}\im\,\xi_Y\to\infty$ removes all $\phi_1$ terms, leaving us with
\begin{equation} \label{Z0Landau}
\begin{aligned}
Z_0\big|_{\textrm{Landau}} 
&\= \tfrac{\im}{3}\,\smallint \vp_{ab}\bigl(\eta^{ab}{-}\tfrac{\pa^a\pa^b}{\Box}\bigr)\,\Box\,\bigl(\eta^{cd}{-}\tfrac{\pa^c\pa^d}{\Box}\bigr)\,\vp_{cd} \\
&\= \tfrac{\im}3\,\smallint \bigl\{ \vp\,\Box\,\eta^{cd} - 2 p\,\vp\,\pa^c\pa^d - 2 q\,\vp_{ab}\pa^a\pa^b\eta^{cd}
+ \vp_{ab}\,\pa^a\pa^b\,\Box^{-1}\pa^c\pa^d \bigr\}\,\vp_{cd} \ .
\end{aligned}
\end{equation}
In the second line we have taken advantage of the freedom to add a total derivative, in order to introduce free parameters
$p$ and~$q$ with $p{+}q=1$.

At this stage we employ the partial Wick theorem for the free correlator $\langle\,Z_0\,Y\,\rangle_0^\vp$.
Noting the permutation symmetry of~\eqref{Z0Landau} and
discarding disconnected terms $\langle\,Z_0\,\rangle_0^\vp\,\langle\,Y\,\rangle_0^\vp$, it suffices to 
convert the last field factor~$\vp_{cd}$ above to a contraction with the Euler operator acting on~$Y$,
\begin{equation}
\vp_{cd} \quad\Rightarrow\quad 
\bcontraction{}{\vp}{_{cd}\quad\smallint}{\vp}\vp_{cd}\quad\smallint\vp_{ab}\,\tfrac{\de}{\de\vp_{ab}} \ ,
\end{equation}
and insert the free graviton propagator~\eqref{boseprop}. The result replaces 
\begin{equation}
Z_0\big|_{\textrm{Landau}}  \ \to\ 
\smallint \vp\,\bigl\{ \tfrac14\,\eta_{ab} + (4p{-}\tfrac32)\,\tfrac{\pa_a\pa_b}{\Box} \bigr\}\,\tfrac{\de}{\de\vp_{ab}}\ .
\end{equation}
The local coefficient~$\tfrac14$ is precisely what is needed.\footnote{
In our earlier work~\cite{AKL} we had cast the $Z$ factor into an $R^{\textrm{mix}}$ contribution but missed some terms along the way.
Therefore, even though the result appeared gauge-independent, it was incomplete, with this coefficient uncancelled.}
Clearly, we can adjust the free parameter~$p$ such as to cancel the gauge-dependent nonlocal term in~\eqref{R0inv},
so that finally
\begin{equation}
\bigl\langle\, R_0^{\textrm{inv}}\,+\,R_0^{\textrm{gf}}\,+\,Z_0\,-\,E \,\bigr\rangle_0^{\vp}\big|_{\textrm{Landau}} \= 0
\end{equation}
as it should be. Therefore, up to gauge artefacts, the flow in the gravitational coupling admits a power series expansion
in~$\ka$ around Minkowski space, and its iteration allows for a perturbative computation of quantum gravity correlators.

\section{Comparison with the on-shell method}\label{sec:onshell}

How does our coupling-flow construction relate to the direct on-shell method initiated in~\cite{DN} and Section~7 of~\cite{AKL}
and recently extended to order~$\ka^2$ in~\cite{CJL}? Diagrammatically, the loop-decorated tree expansion of the Nicolai map
looks as follows,
\newcommand\scale{.70}
\begin{equation} \label{graphs}
\begin{aligned}
T_\kappa\phi\ \=\
\scalebox{\scale}{
\begin{tikzpicture}[baseline={([yshift=-1ex]current bounding box.center)},thick]
\node at (0,0)[circle,fill,inner sep = 2pt] {};
\draw[gluon](1,0)--(0,0);
\end{tikzpicture}
}
 & \ \ +\ \
\kappa \
\scalebox{\scale}{
\begin{tikzpicture}[baseline={([yshift=-1ex]current bounding box.center)},thick]
\node at (0,0)[circle,fill,inner sep = 2pt] {};
\draw (1,0)--(0,0);
\draw[gluon] (1+0.75, 0.5)--(1,0);
\draw[gluon] (1+0.75,-0.5)--(1,0);
\end{tikzpicture}
}
\ \ +\ \ \;
\kappa^2\
\left(
        \scalebox{\scale}{
        \begin{tikzpicture}[baseline={([yshift=-1ex]current bounding box.center)},thick]
        \node at (0,0)[circle,fill,inner sep = 2pt] {};
        \draw (2,0)--(-0,0);
        \draw[gluon] (2+0.75, 0.5)--(2,0);
        \draw[gluon] (2+0.75,-0.5)--(2,0);
        \draw[gluon] (1,-0.9)--(1,0);
        \phantom{ \draw[gluon] (1,0.9)--(1,0); }
        \end{tikzpicture}
        }
        \ \ +\ \
        \scalebox{\scale}{
        \begin{tikzpicture}[baseline={([yshift=-1ex]current bounding box.center)},thick]
        \node at (0,0)[circle,fill,inner sep = 2pt] {};
        \draw (1,0)--(0,0);
        \draw[gluon] (1+0.75, 0.5)--(1,0);
        \draw[gluon] (1+1,0)--(1,0);
        \draw[gluon] (1+0.75,-0.5)--(1,0);
        \end{tikzpicture}
        }
\right)
\ \ \;+\ \ O(\kappa^3) \\
& \ \ +\ \
\hbar\,\kappa \
\scalebox{\scale}{
\begin{tikzpicture}[baseline={([yshift=-1ex]current bounding box.center)},thick]
\node at (0,0)[circle,fill,inner sep = 2pt] {};
\draw (1,0)--(0,0);
\draw [domain=0:360] plot ( {   (  4 +  cos(\x)  )  /3} , {    sin(\x)  /3  } );
\end{tikzpicture}
}
  \ +\
\hbar\,\kappa^2
\left(
        \scalebox{\scale}{
        \begin{tikzpicture}[baseline={([yshift=-1ex]current bounding box.center)},thick]
        \node at (-1,0)[circle,fill,inner sep = 2pt] {};
        \draw (1,0)--(-1,0);
        \draw[gluon] (0,0)--(0,-0.9);
        \phantom{
        \draw[gluon] (0,0)--(0,0.9);
        }
        \draw [domain=0:360] plot ( {   (  4 +  cos(\x)  )  /3} , {    sin(\x)  /3  } );
        \end{tikzpicture}
        }
        \  +\
        \scalebox{\scale}{
        \begin{tikzpicture}[baseline={([yshift=-1ex]current bounding box.center)},thick]
        \node at (0,0)[circle,fill,inner sep = 2pt] {};
        \draw (1,0)--(0,0);
        \draw [domain=0:360] plot ( {   (  4 +  cos(\x)  )  /3} , {    sin(\x)  /3  } );
        \draw[gluon] (1.667+1,0)--(1.667,0);
        \end{tikzpicture}
        }
 \right)
\ \ +\ \ O(\hbar\,\kappa^3) \\
& \qquad\qquad\qquad\qquad\qquad\qquad\qquad\qquad\qquad\qquad\qquad\qquad\qquad\qquad
\ +\ \ O(\hbar^2 \kappa^3) \ ,
\end{aligned}
\end{equation}
where curly lines symbolize vierbein factors, straight lines denote $\Box^{-1}$, and vertices (except for the fat one)
are integrated over. Each propagator line carries two partial derivatives, but each loop reduces their number by two.
The complexity resides in the possible index contractions on partial derivatives and vierbein fields. 
Our minimal solution for $T^{(0,1)}$ below requires just four terms, but the number of terms goes into the hundreds
for $T^{(0,2)}$.\footnote{
In the Feynman gauge $\xi_P{=}1$, \cite{CJL} finds 7 terms for $T^{(0,1)}$, 
294 distinct contributions to $T^{(0,2)}$ and 59 to $T^{(1,2)}$.}
From the diagram we see that $T^{(1,1)}$ is a singular constant, which is normally regularized to zero.
Analyzing the graphical ``Nicolai rules'' leading to~\eqref{graphs} one sees that actually every $r$-loop contribution
starts only at order~$\ka^{2r}$ after a singular constant~$T^{(r,2r-1)}$.
Of course, some index contractions on partial derivatives produce a $\Box$ which collapses a propagator.
This leads to further diagrams not shown here, except for the last one in the first line above, which is of this type.

Here we go to the Landau gauge~$\xi_P{\to}\infty$, which strictly imposes $F_P^b=\pa^b\vp{-}2\pa_a\vp^{ab}=0$,
and consider just the leading $O(\hbar^0\ka^1)$ term~\cite{AKL}, 
\begin{equation} \label{T01}
T^{(0,1)} \vp_{ab}(x) \= \int\!\diff^4y\ \pa^c \pa^d \Box^{-1}(x{-}y)\,\bigl\{
\tfrac12\eta_{cd}\,\vp\,\vp_{ab}\ -\ \tfrac12\eta_{cd}\,\vp_a^{\ e}\vp_{eb}\ -\ \vp_{ab}\vp_{cd}\ +\ 2\,\vp_{ac}\vp_{bd} 
\bigr\}(y)\ ,
\end{equation}
in which only four out of twelve possible index contractions appear.
Four others are excluded by the free-action condition,
and the remaining four are unfixed by the Nicolai-map conditions even to order~$\ka^2$,\footnote{
Hun Jang, private communication. In the Feynman-gauge example of \cite{CJL} there appear
three further terms in~$T^{(0,1)}$, one of which cannot be brought into the form of~\eqref{dT01} 
except on the harmonic-gauge slice~$F_P{=}0$.}
\begin{equation} \label{dT01}
\delta T^{(0,1)} \vp_{ab}(x) \= \int\!\diff^4y\ \pa_{(a} \pa^c \Box^{-1}(x{-}y)\,\bigl\{
\la_1\,\vp\,\vp_{b)c} + \la_2\,\vp_{b)}^{\ e}\vp_{ec} + \la_3\,\eta_{b)c}\,\vp^2 + \la_4\,\eta_{b)c}\,\vp_e^{\ g}\vp_g^{\ e} \bigr\}(y)
\end{equation}
with arbitrary coefficients $\la_1,\ldots,\la_4$, but they change the vierbein only by a longitudinal term.

The defining relation~\eqref{Tdef} for the Nicolai map says nothing about the choice of gauge-fixing needed to render
well-defined either side (for gauge non-invariant functionals~$Y$). For this reason Nicolai originally amended the key properties 
of the map by the requirement that the gauge-fixing function be a fixed point of the map~\cite{Nic2,Nic3}.
Outside the Landau gauge, for $\xi_P<\infty$, the path integral averages over nonzero values of the gauge-fixing function,
and so its vanishing is not strictly imposed, and the map is allowed to shift the gauge slice, starting at order~$\kappa$.
In the Landau gauge, for $\xi_P\to\infty$ however, such a shift appears awkward and is undesirable.
In fact, the flow-operater construction in super Yang--Mills theory automatically produces a map acting within the gauge slice.
For supergravity, the map will respect the harmonic gauge fixing if
\begin{equation}
T_\ka F^b_P[\vp_{\cdot\cdot}] \= F^b_P[T_\ka\vp_{\cdot\cdot}] \= \pa^b T_\ka\vp - 2\pa_a T_\ka\vp^{ab} \= 0
\qquad\for\quad \pa^b\vp - 2\pa_a\vp^{ab} = 0\ ,
\end{equation}
and the linearity of the gauge condition implies this condition to hold separately for each component~$T^{r,n}$.
However, it is easy to see that
the above expressions for $T^{(0,1)}$ and $\de T^{(0,1)}$ do not respect the harmonic gauge condition.
Yet, this is easily repaired by applying the harmonic-gauge projector
\begin{equation}
P_{ab}^{\ \ cd} \= 
\tfrac12(\de_a^c{-}\tfrac{\pa_a\pa^c}{\Box})(\de_b^d{-}\tfrac{\pa_b\pa^d}{\Box}) +
\tfrac12(\de_a^d{-}\tfrac{\pa_a\pa^d}{\Box})(\de_b^c{-}\tfrac{\pa_b\pa^c}{\Box}) +
\tfrac{\pa_a\pa_b}{\Box}(\eta^{cd}{-}\tfrac{\pa^c\pa^d}{\Box})\ .
\end{equation}
Therefore, a Nicolai map strictly respecting the harmonic gauge starts out with
\begin{equation}
\begin{aligned}
T^{(0,1)}_{\textrm{Landau}} \vp_{ab}(x) \ :=\ & P_{ab}^{\ \ cd}\,T^{(0,1)}\vp_{cd}(x) \\[2pt]
\= T^{(0,1)} \vp_{ab}(x)  &\ +
\int\!\diff^4y\ \pa_{(a} \pa^c \Box^{-1}(x{-}y)\,\bigl\{
\vp\,\vp_{b)c} - \vp_{b)}^{\ e}\vp_{ec} -\tfrac12\,\eta_{b)c}\,\vp^2 + 
\tfrac12\,\eta_{b)c}\,\vp_e^{\ g}\vp_g^{\ e} \bigr\}(y) \\ &\ +
\int\!\diff^4y\  \pa_{(a}\pa^c\pa^d\pa^e \Box^{-2}(x{-}y)\,\bigl\{
2\,\vp_{b)c}\vp_{de} + \eta_{b)e}\,\vp\,\vp_{cd} -2\,\eta_{b)e}\,\vp_c^{\ g}\vp_{gd} \big\}(y)\ .
\end{aligned}
\end{equation}
This adds a specific contribution to $\de T^{(0,1)}$ in~\eqref{dT01} but also generates three higher nonlocal terms
not included in the ansatz of~\cite{CJL}, 
which picked the Feynman gauge and does not respect the gauge condition. 
We note that the additional terms are all of gradient type and thus 
lie in the kernel of the projector, $P_{ab}^{\ \ cd}\pa_cX_d=0$.

Given a Nicolai map~$T_\ka$, it is straightforward to write down documenta corresponding flow operator
\begin{equation}
R[\vp^\cdot_{\ \cdot};\ka] \= \int\!\diff^4y\ \bigl( \ka\pa_\ka T_\ka^{-1}\circ T_\ka\bigr)\,\vp^a_{\ \mu}(y)\,
\frac{\de}{\de\vp^a_{\ \mu}(y)}\ ,
\end{equation}
but such an operator is only beneficial when it can be constructed independently thanks to supersymmetry.
In all known cases, $R$ is found as an average over the integrated-out fields~$\phi_i$ of the product of a
fermionic functional~$\Delt$ with a supervariation, possibly amended by a Slavnov variation.
Knowing the Nicolai map to leading order, we can thus read off a leading-order flow operator and
investigate whether it can be reverse-engineered to such a form.
To streamline the notation we abbreviate
\begin{equation}
\vp{\cdot}\vp_{ab} := \vp_a^{\ e}\vp_{eb}  \quad\und\quad \vp{:}\vp := \vp_e^{\ g}\vp_g^{\ e} \ .
\end{equation}
From \eqref{Tinvformula} and~\eqref{Rgexp} one simply reads off that
\begin{equation} \label{R1onshell}
R_1 \= -\int\!\diff^4y\int\!\diff^4x\ \bigl\{
\tfrac12\,\vp\,\vp_{ab}\,\eta_{cd} - \tfrac12\,\vp{\cdot}\vp_{ab}\,\eta_{cd} - \vp_{ab}\vp_{cd} + 2\,\vp_{ac}\vp_{bd} 
\bigr\}(y)\ \pa^c \pa^d \Box^{-1}(y{-}x)\ \frac{\de}{\de\vp_{ab}(x)}
\end{equation}
in leading order $\hbar^0$ and ignoring the $\de T$ ambiguity above.

Leaving out the ghosts for the moment, the coupling-flow method requires the flow operator in leading order to take the form
\begin{equation} \label{R1ansatz}
R_1 \= \im\,\smallint\bcontraction{}{\Delt}{^{(1)}_\al\ \smallint}{\de}\Delt^{(1)}_\al\ \smallint\de_\al\vp_{ab}\,\tfrac{\de}{\de\vp_{ab}}
\end{equation}
with a leading-order fermionic functional
\begin{equation} \label{M1ansatz}
\Delt^{(1)}_\al \= \vp^{ab}\vp^{cd}\,(N^{\mu\nu}_{ab,cd})_{\al\be}\,(\pa_\mu\psi_\nu)_\be 
\end{equation}
containing a coefficient matrix~$N$ built entirely from gamma matrices (and the metric). 
Modulo obvious symmetries, it produces eleven different index contractions. Using
\begin{equation}
\de_\al \vp_{ab} \= -\tfrac12 (\bar\psi_b\ga_a)_\al \quad\und\quad
\bcontraction{}{\psi}{_c\quad}{\bar\psi}\psi_c\quad\bar\psi_d \= \tfrac{3\im}{2}\ga_d\tfrac{\slashed\pa}{\Box}\ga_c
\end{equation}
it is straightforward to evaluate the gamma traces.
With the unique combinations
\begin{equation}
\begin{aligned}
\hat{\Delt}^{(1)} &\=  -\tfrac1{24}\bigl\{ [\vp^2-\vp{:}\vp]\ga^\nu\ga^\mu
+2[\vp\,\vp^\mu_a-\vp{\cdot}\vp^\mu_a]\ga^{a\nu}-2[\vp\,\vp^\nu_a-\vp{\cdot}\vp^\nu_a]\ga^{a\mu} 
\bigr\}\,\pa_\mu\psi_\nu\ ,\\
\check{\Delt}^{(1)} &\=  -\tfrac18\bigl\{ -[\vp\,\vp^{\mu\nu}-\vp{\cdot}\vp^{\mu\nu}]
+ [\vp\,\vp^\mu_a-\vp{\cdot}\vp^\mu_a]\ga^{a\nu} - 2\,\vp_a^\mu\vp_b^\nu\,\ga^{ab}  
\bigr\}\,\pa_\mu\psi_\nu
\end{aligned}
\end{equation}
one generates precisely the first two and the last two terms in~\eqref{R1onshell}, respectively,
\begin{equation}
\begin{aligned}
\hat{R}_1 &\= -\smallint\!\smallint \bigl\{ \tfrac12\,\vp\,\vp_{ab} - \tfrac12\,\vp{\cdot}\vp_{ab} \bigr\}\,\eta_{cd}\
\pa^c \pa^d \Box^{-1}\ \tfrac{\de}{\de\vp_{ab}}\ ,\\
\check{R}_1 &\= -\smallint\!\smallint \bigl\{ -\tfrac32\,\vp_{ab}\vp_{cd} + \tfrac32\,\vp_{ac}\vp_{bd} \bigr\}\ 
\pa^c \pa^d \Box^{-1}\ \tfrac{\de}{\de\vp_{ab}}\ ,
\end{aligned}
\end{equation}
but the relative coefficients are necessarily equal and opposite in sign, which contradicts 
the coefficients (-1,2) in~\eqref{R1onshell}. As already remarked in~\cite{AKL}, 
the two terms in~$\check{R}_1$ arise purely from the last term in~$\check{\Delt}^{(1)}$,
which produces only an antisymmetric combination. Hence, the missing symmetric part,
\begin{equation}
\bar{R}_1 \= -\smallint\!\smallint \bigl\{ \tfrac12\,\vp_{ab}\vp_{cd} + \tfrac12\,\vp_{ac}\vp_{bd} \bigr\}\ 
\pa^c \pa^d \Box^{-1}\ \tfrac{\de}{\de\vp_{ab}}\ ,
\end{equation}
cannot be obtained from a flow operator of type~\eqref{R1ansatz} with~\eqref{M1ansatz}.
Extending the ansatz~\eqref{R1ansatz} by ghost terms of the form
\begin{equation}
R_1^{\textrm{gf}} \= \im\,\smallint\bcontraction{}{\Delt}{^{(1)}_{\textrm{gf}}\ \smallint}{s}
\Delt^{(1)}_{\textrm{gf}}\ \smallint s\,\vp_{ab}\,\tfrac{\de}{\de\vp_{ab}}
\end{equation}
does not alleviate this problem. However, a symmetric term of the form~$\bar{R}_1$ {\it can\/} 
arise from a partial Wick contraction in the final free graviton correlator, because the free graviton
propagator is symmetric. This suggests again that the flow-operator method can only be 
salvaged by such a procedure.

\section{Conclusions}\label{sec:outro}

We have established the all-order existence of a flow in the gravitational coupling~$\ka$ for Poincar\'e supergravity.
Off-shell global supersymmetry in the superconformal formulation suffices to generate the gauge-invariant part 
of the flow for the effective graviton theory after integrating out all other field degrees of freedom. 
Due to the non-commutativity of supersymmetry and gauge transformations, the BRST gauge fixing
produces multiplicative terms in the flow equation, which obstructs the direct connection with the (inverse)
Nicolai map. We found evidence, however, that this obstacle might be overcome with the help of Wick's theorem
in the free-field correlator. Regardless, the coupling flow may be employed in Poincar\'e supergravity for an 
alternative perturbation theory around Minkowski space. In the Landau gauge and to order~$\ka$, the Nicolai map
requires only four terms, with an ambiguity of four more. Its all-order existence is still open.
It will also be interesting to apply the present method to the simpler cases of three- and two-dimensional supergravity
(the latter being a topological theory). These theories may provide useful testing grounds for the Wick-contraction 
argument developed here. It remains to be seen whether their superconformal versions~\cite{KT,BKNT,KR} are helpful
in this context as well.

\subsection*{Acknowledgments}
We thank Hun Jang for illuminating discussions and a generous appraisal of our previous work.
Insightful discussions with Hermann Nicolai are also acknowledged.
%



\end{document}